\documentclass[iop]{emulateapj}
\pdfoutput=1
\usepackage{apjfonts}
\usepackage[breaklinks,colorlinks,urlcolor=blue,citecolor=blue,linkcolor=blue]{hyperref}
\journalinfo{The Astrophysical Journal, 2015, in press}
\usepackage{amsmath}
\usepackage{natbib}
\usepackage{color}
\definecolor{changes}{rgb}{0.2,0.6,0.2}
\begin{document}

\title{Time-Dependent Turbulent Heating of Open Flux Tubes in the Chromosphere, Corona, and Solar Wind}
\author{L. N. Woolsey\altaffilmark{1} and S. R. Cranmer\altaffilmark{1,2}}
\email{lwoolsey@cfa.harvard.edu}

\altaffiltext{1}{Harvard-Smithsonian Center for Astrophysics, 60 Garden St., Cambridge, MA 02138, USA}
\altaffiltext{2}{Department of Astrophysical \& Planetary Sciences, University of Colorado, Boulder, CO 80309, USA}

\begin{abstract}
We investigate several key questions of plasma heating in open-field regions of the corona that connect to the solar wind. We present results for a model of Alfv\'{e}n-wave-driven turbulence for three typical open magnetic field structures: a polar coronal hole, an open flux tube neighboring an equatorial streamer, and an open flux tube near a strong-field active region. We compare time-steady, one-dimensional turbulent heating models \citep{2007ApJS..171..520C} against fully time-dependent three-dimensional reduced-magnetohydrodynamics modeling of BRAID \citep{2011ApJ...736....3V}. We find that the time-steady results agree well with time-averaged results from BRAID. The time-dependence allows us to investigate the variability of the magnetic fluctuations and of the heating in the corona. The high-frequency tail of the power spectrum of fluctuations forms a power law whose exponent varies with height, and we discuss the possible physical explanation for this behavior. The variability in the heating rate is bursty and nanoflare-like in nature, and we analyze the amount of energy lost via dissipative heating in transient events throughout the simulation. The average energy in these events is $10^{21.91}$ erg, within the ``picoflare'' range, and many events reach classical ``nanoflare'' energies. We also estimated the multithermal distribution of temperatures that
would result from the heating-rate variability, and found good agreement with observed widths of coronal differential emission measure (DEM) distributions. The results of the modeling presented in this paper provide compelling evidence that turbulent heating in the solar atmosphere by Alfv\'{e}n waves accelerates the solar wind in open flux tubes.\\
\end{abstract}

\section{Introduction}
The solar atmosphere is characterized by a temperature profile that has puzzled scientists for decades. The photosphere, the visible surface of the Sun, is roughly 6000 K, while the diffuse corona above it is heated to millions of degrees. Observations have not been able to determine the mechanism(s) responsible for the extreme coronal heating, though many physical processes have been suggested. The same processes that might explain the temperature of the corona can also account for the acceleration of the solar wind. Only a small fraction of the mechanical energy in the Sun's sub-photospheric convection zone needs to be converted to heat in order to power the corona. However, it has proved exceedingly difficult to distinguish between competing theoretical models using existing observations. Recent summaries of these problems and controversies have been presented by, e.g., \citet{2012RSPTA.370.3217P}, \citet{2012SSRv..172...69M}, \citet{2014arXiv1410.5660K}, and \citet{2014arXiv1412.2307C}.

If solar wind flux tubes are open to interplanetary space, and if they remain open on timescales comparable to the time it takes plasma to accelerate into the corona, then the main sources of energy must be injected at the footpoints of the flux tubes. Thus, in wave/turbulence-driven models, the convection-driven jostling of the flux-tube footpoints is assumed to generate wave-like fluctuations that propagate up into the extended corona. These waves (usually Alfv\'en waves) are often proposed to partially reflect back down toward the Sun, develop into strong magnetohydrodynamic (MHD) turbulence, and dissipate gradually. In this project, we use two models in this regime, ZEPHYR \citep{2007ApJS..171..520C} and BRAID \citep{2011ApJ...736....3V}. These and other such models have been shown to naturally produce realistic fast and slow winds with wave amplitudes of the same order of magnitude as those observed in the corona and heliosphere \citep[see also][]{1986JGR....91.4111H, 1999ApJ...523L..93M, 2006JGRA..111.6101S, 2010LRSP....7....4O, 2010ApJ...708L.116V, 2011ApJ...743..197C, 2014ApJ...784..120L, 2014MNRAS.440..971M, 2014ApJ...787..160W}. 

However, between the above ideas and others that claim reconnection plays the primary role in heating, nothing has been ruled out because (1) we have not yet made the observations needed to distinguish between the two paradigms, and (2) most models still employ free parameters that can be adjusted to improve the agreement with existing observational constraints. A complete solution must account for all important sources of mass and energy into the three-dimensional and time-varying corona. Regardless of whether the dominant coronal fluctuations are wave-like or reconnection-driven, they are generated at small spatial scales in the lower atmosphere and are magnified and ``stretched'' as they propagate up in height. Their impact on the solar wind's energy budget depends crucially on the multi-scale topological structure of the Sun's magnetic field. 

The current generation of time-steady 1D turbulence-driven solar wind models \citep[e.g.][]{2007ApJS..171..520C, 2010ApJ...708L.116V, 2011ApJ...743..197C, 2013ApJ...767..125C,2014ApJ...784..120L} contain detailed descriptions of many physical processes relevant to coronal heating and solar wind acceleration. However, none of them self-consistently simulate the actual process of MHD turbulent cascade as a consequence of the partial reflections and nonlinear interactions of Alfv\'enic wave packets. Thus, it is important to improve our understanding of the flux-tube geometry of the open-field corona and how various types of fluctuations interact with those structures by using higher-dimensional and time-dependent models to understand the underlying physical processes.

In Section 2, we describe the modeling used to improve our understanding of the geometry of flux tubes in a time-steady sense and extending such understanding to time-dependent three-dimensional modeling. We focus on three models that are typical of common structures in the solar corona and present results of each in Section 3 and compare the one-dimensional, time-steady results from earlier work using ZEPHYR with the new three-dimensional, time-dependent results. We then analyze the polar coronal hole BRAID model further in Section 4. We include a discussion and conclusions in Section 5.

\section{MHD Turbulence Simulation Methods}
\subsection{Previous time-steady modeling}
\begin{figure*}[!ht]
\includegraphics[width=\textwidth]{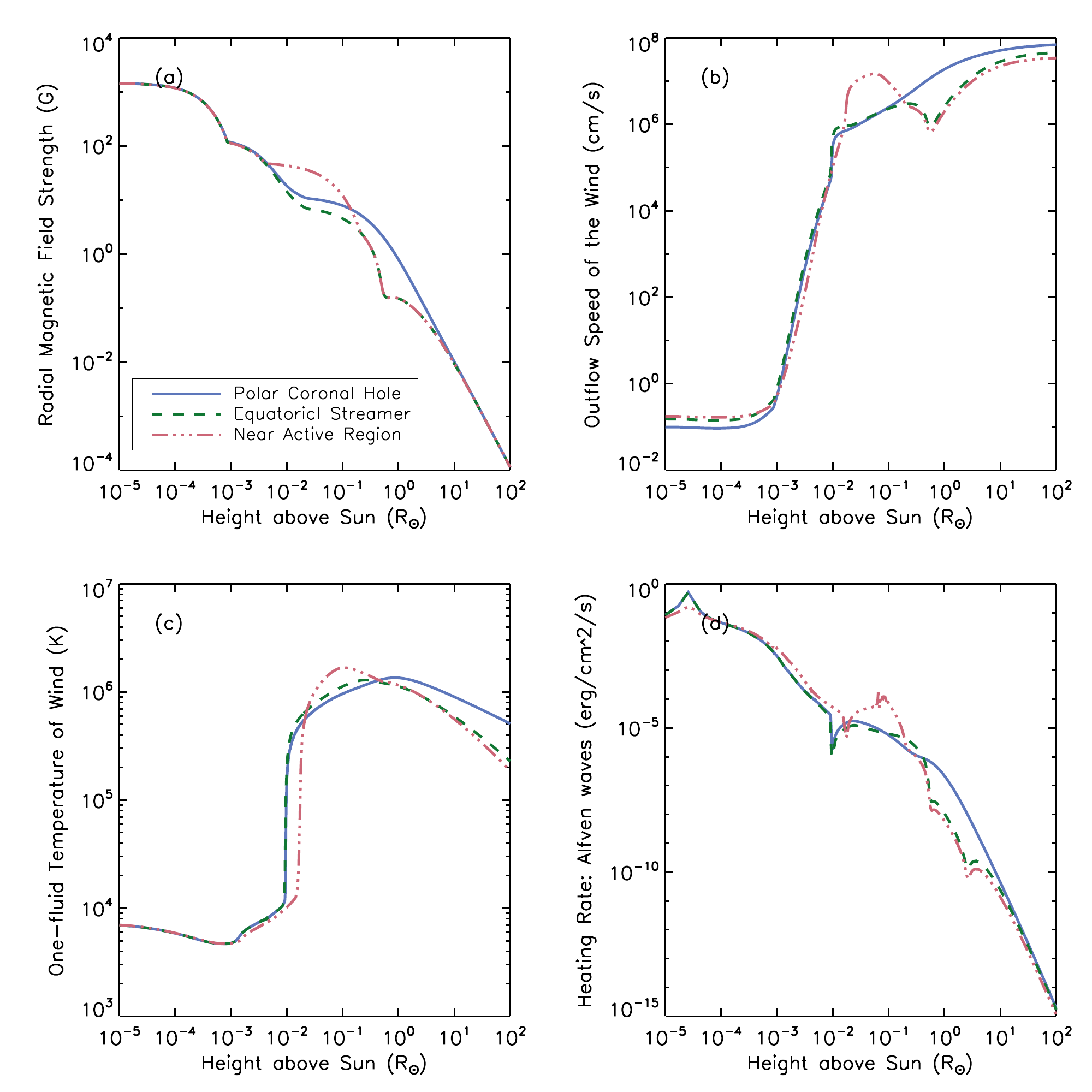}
\caption{The primary input to ZEPHYR are (a) the magnetic profiles of three representative coronal structures. Results for the bulk solar wind properties from each flux tube include: (b) outflow speed, (c) one-fluid temperature, and (d) Alfv\'enic heating rate.}
\label{fig:zephyr3}
\end{figure*}

The magnetic field profiles used for this project are listed in Table \ref{tab:threemodels} and are representative of a polar coronal hole, an equatorial streamer, and a flux tube neighboring an active region. The radial profiles of magnetic field strength $B(r)$ for these models are shown in Figure \ref{fig:zephyr3}a \citep[see also][]{1998A&A...337..940B,2007ApJS..171..520C}. The Alfv\'en travel time from the base of the model to a height of 2 solar radii is calculated using Equation \eqref{eq:traveltime}, and the expansion factor is defined by \citet{1990ApJ...355..726W}. The expansion factor is a ratio of the magnetic field strength at a height of 1.5 solar radii to the field strength at the photospheric base (here, at a height of 0.04 solar radii, the size of a supergranule), normalized such that an expansion factor of 1 is simple radial expansion, and an expansion factor larger than 1 is considered ``superradial'' expansion. Key results from these three models using ZEPHYR, a one-dimensional time-steady code introduced by \citet{2007ApJS..171..520C}, are presented in Figure \ref{fig:zephyr3}.

\begin{table*}[!ht]
    \caption {Models in BRAID}
    \begin{tabular}{|c|c|c|c|c|}
    \hline
    {\bf Modeled structure}  & {\bf Travel time to 2 R$_{\odot}$} & {\bf Expansion factor} & {\bf Speed at 1 AU} & {\bf Identifier} \\ \hline \hline
    Polar Coronal Hole & 770 s & 4.5 & 720 km s$^{-1}$ & PCH \\ \hline
    Equatorial Streamer & 3550 s & 9.1 & 480 km s$^{-1}$ & EQS \\ \hline
    Near Active Region & 4400 s & 41 & 450 km s$^{-1}$ & NAR  \\ \hline
    \end{tabular}
    \label{tab:threemodels}
\end{table*}

The polar coronal hole model has the smallest expansion from the photosphere to the low corona, and the highest wind speed at 1 AU. This relationship was first seen in empirical fits to observations \citep[see, e.g.,][]{1990ApJ...355..726W,2000JGR...10510465A} and has also been seen in models \citep[see, e.g.,][]{2014ApJ...787..160W}. The equatorial streamer model represents an open flux tube directly neighboring the helmet streamers seen around the equator at solar minimum, and produces slower wind at 1 AU. The flux tube neighboring an active region has a stronger magnetic field above the transition region and an even slower wind speed at 1 AU than the equatorial streamer model. In these models, the transition region for PCH and EQS is at a height of 0.01 solar radii, and the NAR model has a transition region slightly higher, at a height of 0.015 solar radii. The density profiles were determined self-consistently with the wind speed using mass flux conservation.

ZEPHYR solves for a steady state solution to the solar wind properties generated by a one-dimensional open flux tube. By solving the equations of mass, momentum, and energy conservation and iterating to a stable solution, the code produces solar wind with mean properties that match observations and {\it in situ} measurements \citep{2007ApJS..171..520C,2014ApJ...787..160W}. The code's expression for the turbulent heating is a phenomenological cascade rate whose form has been guided and validated by several generations of numerical simulations and other models of imbalanced, reflection-driven turbulence \citep[see, e.g.,][]{1995PhFl....7.2886H,2007ApJ...655..269L,2009ApJ...701..652C,oughton15}. For additional details, see Section 3.2 below.

ZEPHYR can only take us so far, however. It is only by modeling the fully 3D spatial and time dependence of the cascade process (together with the intermittent development of magnetic islands and current sheets on small scales) that we can better understand the way in which the plasma is heated by the dissipation of turbulence. We therefore make use of the time-dependent modeling of coronal turbulence introduced by \citet{2011ApJ...736....3V} using the reduced magnetohydrodynamics (RMHD) code called BRAID.

\subsection{Including time-dependence and higher dimensions}
Previous numerical simulations of reflection-driven RMHD turbulence with the full nonlinear terms include \citet{2003ApJ...597.1097D}, \citet{2013ApJ...776..124P}, and prior studies using BRAID on closed loops \citep{2011ApJ...736....3V,2012ApJ...746...81A,2013ApJ...773..111A,2014ApJ...786...28A}. Our version of BRAID uses the three-dimensional equations of RMHD \citep{1976PhFl...19..134S,1982PhST....2...83M,1992JPlPh..48...85Z,1998ApJ...494..409B} to solve for the nonlinear reactions between Alfv\'en waves generated at the single footpoint of an open flux tube. RMHD relies on the assumption that the incompressible magnetic fluctuations $\delta{\bf B}$ in the system are small compared to an overall background field ${\bf B}_{0}$. At scales within the turbulence inertial range, observations show that $\delta{\bf B}$ is much smaller in amplitude than the strength of the surrounding magnetic field ${\bf B}$ and $\delta{\bf B}$ is perpendicular to ${\bf B}$ \citep{1990JGR....9520673M,1995SSRv...73....1T,2012ApJ...758..120C}. 

Some implementations of RMHD combine together the implicit assumptions of incompressible fluctuations, high magnetic pressure (i.e., plasma $\beta \ll 1$), and a uniform background field ${\bf B}_0$. However, the flux tubes we model in the upper chromosphere and low corona have some regions with $\beta \approx 1$ and a vertical field ${\bf B}_0$ that declines rapidly with height. It is not necessarily the case that RMHD applies in this situation. However, it was shown in Section 3.1 of \citet{2011ApJ...736....3V} that there is a self-consistent small-parameter expansion that gives rise to a set of RMHD equations appropriate for the chromosphere and corona. In these equations, the dominant, first-order fluctuations are transverse and incompressible---even when $\beta \approx 1$---and gravitational stratification is included to account for the height variation of ${\bf B}_0$.

BRAID uses two cross-sectional dimensions of the flux tube and a third dimension along the length of the flux tube, which aligns with the background field $B_{0}$. Alfv\'en waves are generated at the lower boundary by random footpoint motions with an rms velocity of 1.5 km s$^{-1}$ and correlation time of 60 s. These fluctuations are generated by taking a randomized white-noise time stream and passing it through a low-pass (Gaussian) frequency filter that removes fluctuations shorter than the specified correlation time. The time stream is normalized to the desired rms velocity amplitude and split up between two orthogonal low-$k_{\perp}$ Bessel-function modes of the cylindrically symmetric system. The driver modes are shown in Figure \ref{fig:sincos} and are discussed further in Appendix B of \citet{2011ApJ...736....3V}.

\begin{figure}[!ht]
\includegraphics[width=\columnwidth]{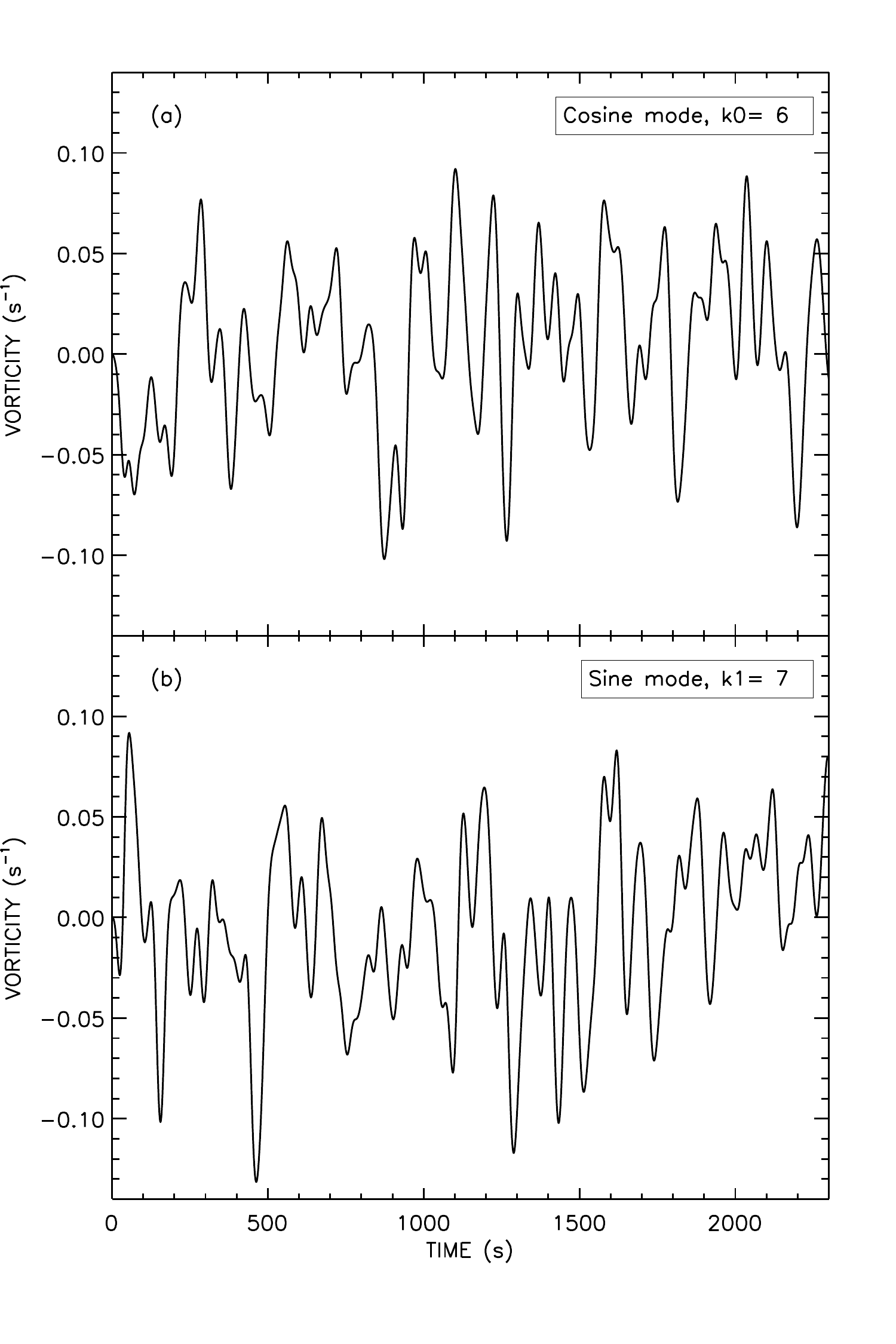}
\caption{We show here vorticity $\omega$ from (a) cosine and (b) sine driver modes for the PCH.}
\label{fig:sincos}
\end{figure}

The magnetic and velocity fluctuations can be approximated by 
\begin{subequations}
\begin{align}
{\bf \delta B} = \nabla_{\perp}h \times {\bf B},\\
{\bf \delta v} = \nabla_{\perp}f \times {\bf \hat{B}},
\end{align}
\end{subequations}
\noindent where ${\bf B}$ is the full magnetic field vector, whose magnitude varies with height, and $\hat{\bf B}$ is the unit vector along the magnetic field. Also, $h({\bf r},t)$ is a height- and time-dependent function that is analogous to the standard RMHD magnetic flux function, and $f({\bf r},t)$ is a velocity stream function (sometimes called $\psi$ in other derivations of RMHD). We can also define the magnetic torsion parameter $\alpha \equiv -\nabla_{\perp}^{2}h$ and the parallel component of vorticity $\omega \equiv -\nabla_{\perp}^{2}f$ \citep[see, e.g.,][]{1982PhST....2...83M}. The functions $h({\bf r},t)$ and $f({\bf r},t)$ satisfy the coupled equations
\begin{subequations}\begin{small}
\begin{align}
\frac{\partial{\omega}}{\partial{t}} + {\bf \hat{B}} \cdot \left(\nabla_{\perp}\omega \times \nabla_{\perp}f\right) = \nonumber\\v_{\rm A}^{2}\left[{\bf \hat{B}} \cdot \nabla{\alpha} + {\bf \hat{B}} \cdot \left(\nabla_{\perp}\alpha \times \nabla_{\perp}h\right)\right] + D_{v},\\
\frac{\partial{h}}{\partial{t}} = {\bf \hat{B}} \cdot \nabla{f} + \frac{f}{H_{B}} + {\bf \hat{B}} \cdot \left(\nabla_{\perp}f \times \nabla_{\perp}h\right) + D_{m},
\end{align}\end{small}
\end{subequations}
where $H_{B}$ is the magnetic scale length and $D_{v}$ and $D_{m}$ describe the effects of viscosity and resistivity \citep[see detailed derivation in][]{2011ApJ...736....3V}. Both $D_v$ and $D_m$ have a hyperdiffusive $k_{\perp}^{4}$ dependence so that the smallest eddies are damped preferentially and the cascade is allowed to proceed without significant damping over most of the $k_{\perp}$ inertial range.

In this paper, we have extended previous work using BRAID by using an open upper boundary condition instead of a closed coronal loop. The model extends to a height of $z_{\rm top} = 2 R_{\odot}$. This height was chosen to model as much of the solar wind acceleration region as possible, without extending into regions where the wind speed becomes an appreciable fraction of the Alfv\'en speed, since the RMHD equations of BRAID don't include the outflow speed. To implement the open upper boundary condition, we set $\omega_{+}$ (which corresponds to the downward Els\"asser variable $Z_{+}$) to zero, whereas in the previous coronal loop models that used BRAID, it was set to the opposite footpoint's boundary time stream as described above \citep[][and references therein]{2014ApJ...786...28A}.

\section{Time-Averaged Results from Three Modeled Flux Tubes}
\begin{figure}[!ht]
\includegraphics[width=\columnwidth]{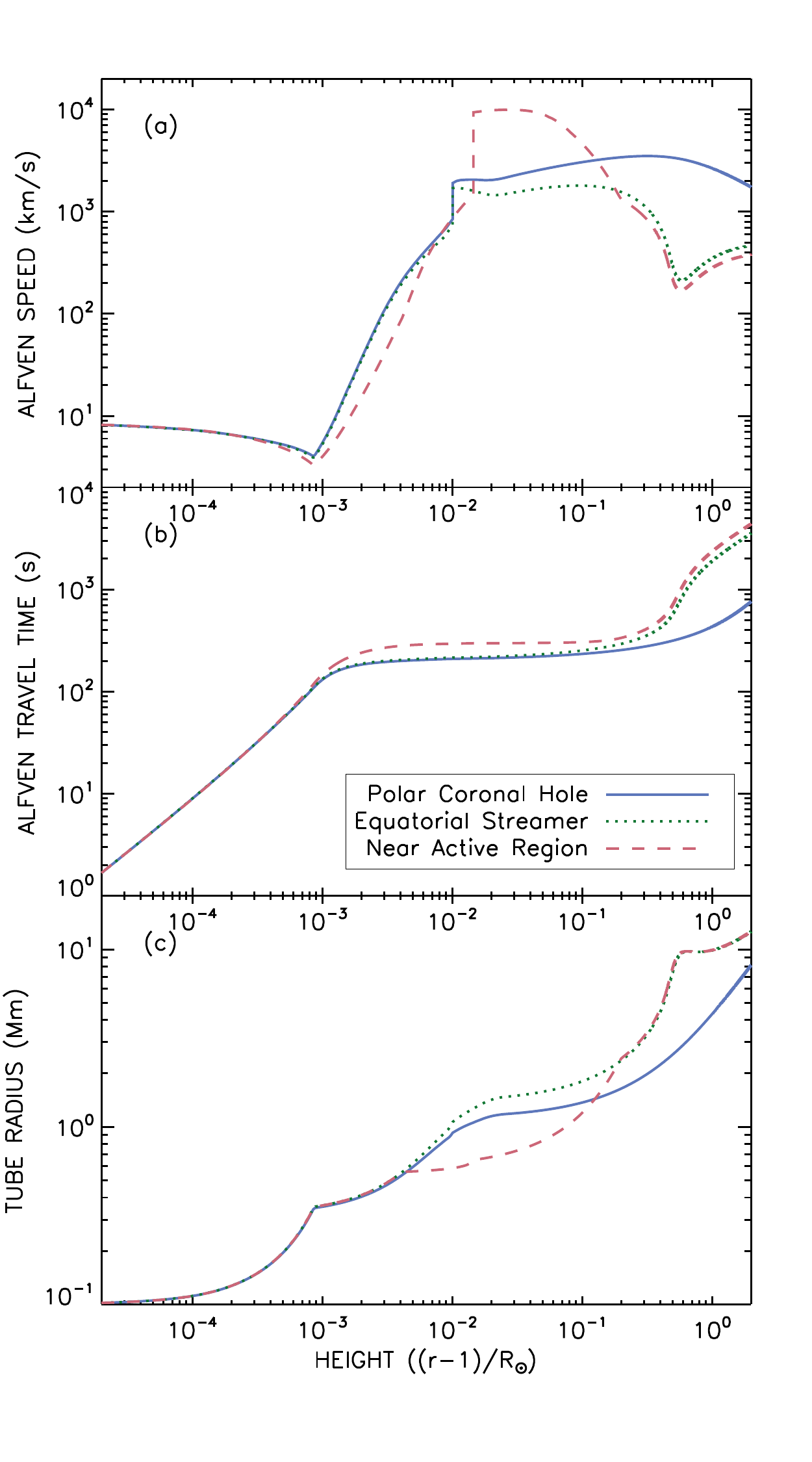}
\caption{Radial profiles for (a) Alfv\'en speed, (b) Alfv\'en travel time, and (c) tube radius.}
\label{fig:models}
\end{figure}

In this paper, we present three models that are representative of common coronal structures (the same from Table \ref{tab:threemodels}). Figure \ref{fig:models} gives some of the time-steady background variables for the three models, based on the input magnetic field profiles shown in Figure \ref{fig:zephyr3}. 

The Alfv\'{e}n speed, $V_{\rm A} = B / \sqrt{4\pi\rho}$ shows a rapid rise in the upper chromosphere, followed by varied behavior in the corona depending on the model. The Alfv\'{e}n travel time is defined as a monotonically increasing function of height,
\begin{equation}\tau_{\rm A} (z) \, = \, \int_{0}^{z} \frac{dz'}{V_{\rm A}(z')}\label{eq:traveltime}\end{equation} where $z=0$ is the photospheric lower boundary of each model. The BRAID code uses $\tau_{\rm A}$ as the primary height coordinate. Figure \ref{fig:models} shows the modeled transverse radius of the flux tube, which is normalized to 100~km at $z=0$ (i.e., a typical length scale for an intergranular bright point) and is assumed to remain proportional to $B^{-1/2}$ in accordance with magnetic flux conservation.

\begin{figure*}[!ht]
\includegraphics[width=\textwidth]{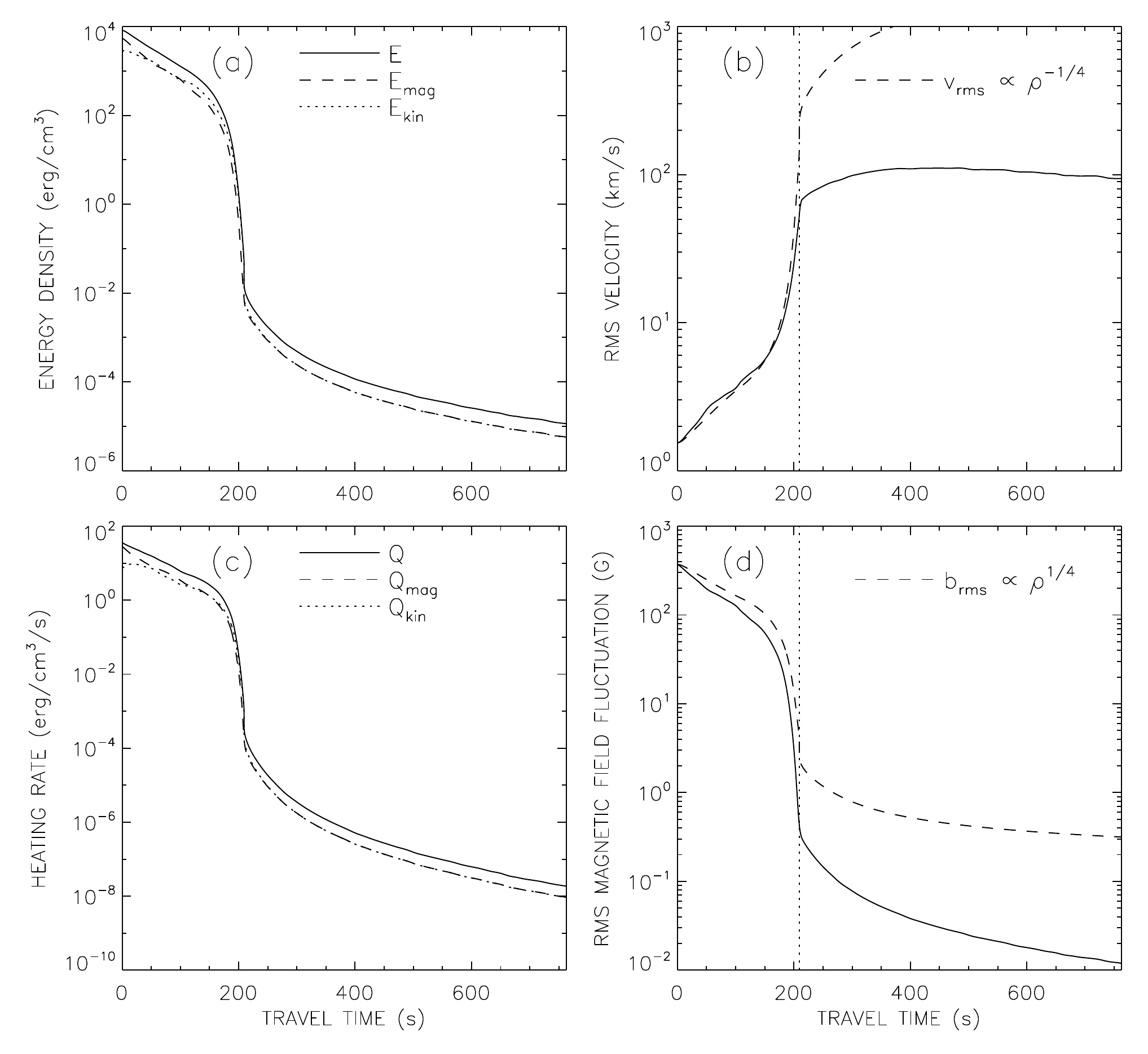}
\caption{The time-averaged results for the PCH model are shown as a function of Alfv\'en travel time as a proxy for height above photosphere: (a) Energy density, (b) rms velocity, (c) heating rate, and (d) rms magnetic field fluctuation.}
\label{fig:pch}
\end{figure*}

Figure \ref{fig:pch} provides the time-averaged results from BRAID for the PCH. The transition region is at a travel time of roughly 210 s and is shown with a dotted line. In Figure \ref{fig:pch}a, we show the magnetic, kinetic, and total energy densities of the
RMHD fluctuations. We plot these quantities separately since no equipartition is assumed. Figure \ref{fig:pch}b shows the increase in the rms transverse velocity amplitude, $\Delta v_{\rm rms}$ with increasing height, which roughly follows the expected sub-Alfv\'enic WKB relation $\Delta v_{\rm rms} \propto \rho_{0}^{-1/4}$ below the transition region. The heating rate is also broken up into magnetic (from the $D_m$ term), kinetic (from the $D_v$ term), and total for Figure \ref{fig:pch}c. Finally, we show the magnetic field fluctuation as a function of travel time in Figure \ref{fig:pch}d. Like the rms velocity, the sub-Alfv\'enic WKB relation $\Delta b_{\rm rms} \propto \rho_{0}^{1/4}$ is followed below the transition region but diverges above it.

\begin{figure}[!h]
\includegraphics[width=\columnwidth]{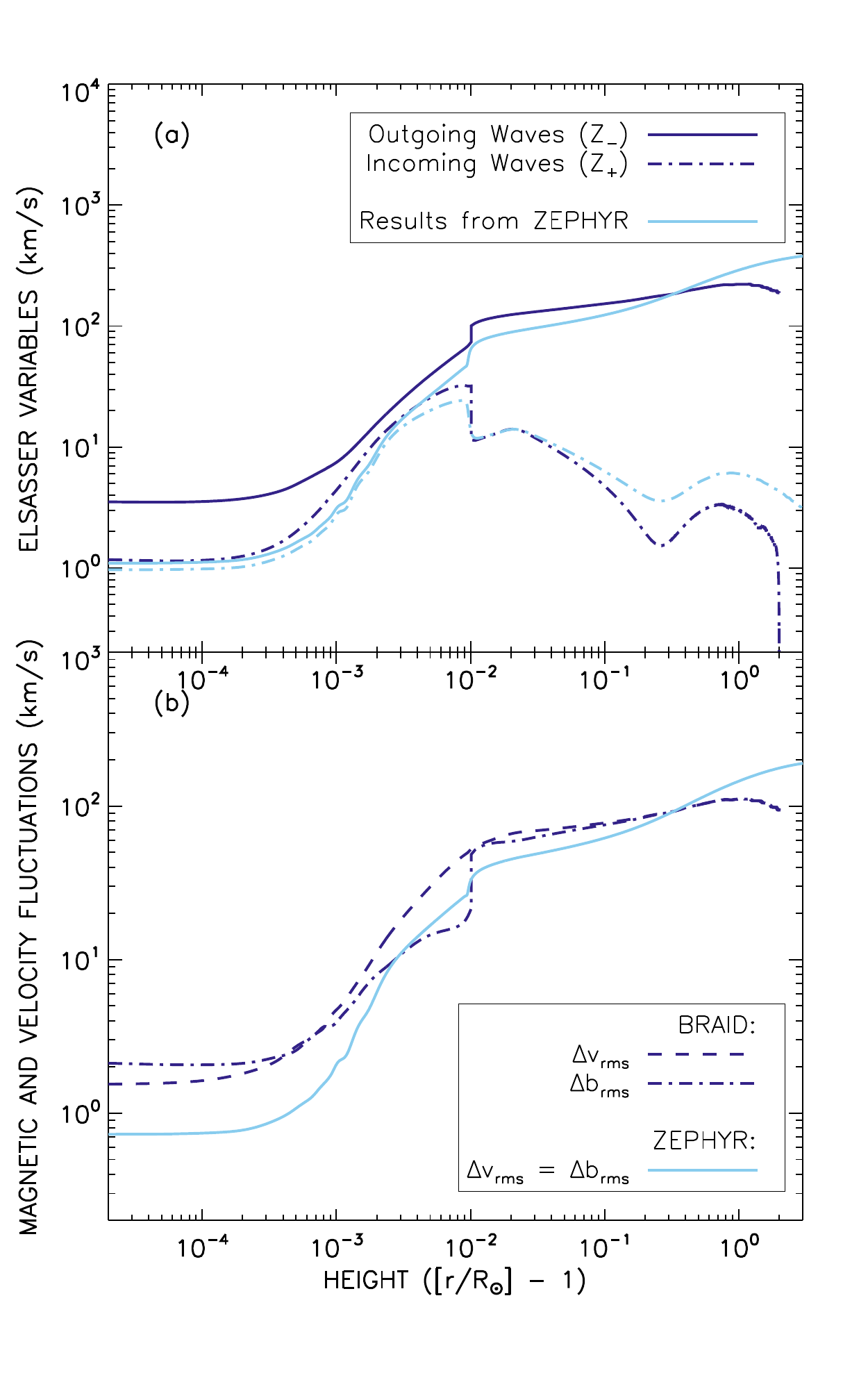}
\caption{For the PCH model, we show (a) the time-averaged amplitudes of incoming and outgoing Alfv\'en waves and (b) the rms amplitudes of magnetic and velocity fluctuations from BRAID and ZEPHYR.}
\label{fig:elsa}
\end{figure}

The quantities plotted in Figures \ref{fig:pch}b, \ref{fig:pch}d, and \ref{fig:elsa}b have gone through two ``levels'' of root mean square averaging. First, we take the variance over all $k_{\perp}$ modes at a given height and time, then take the square root.  Then, at each height, we take an average over the simulation-time dimension using the squares of the first level of rms values. It is those quantities that we then take the square root of and plot.

In Figure \ref{fig:elsa}a, we show the magnitude of the Els\"asser variables, ${\bf Z_{\pm}} = {\bf \delta v} \pm {\bf \delta B} / \sqrt{4\pi \rho_0}$, for the PCH model, and show the fluctuations $\Delta v_{\rm rms}$ and $\Delta b_{\rm rms}$ in Figure \ref{fig:elsa}b. Note that here $\Delta b_{\rm rms}$ is in velocity units, as the actual fluctuations are divided by $\sqrt{4 \pi \rho_{0}}$ (where $\rho_{0}$ is the background density) for comparison with the velocity fluctuations. Our boundary condition enforces that the incoming waves ($Z_{+}$) have exactly zero amplitude at the upper boundary of our model. Figure \ref{fig:elsa}a also shows the radial dependence of the Els\"asser variables from the ZEPHYR model for this coronal hole presented by \citet{2007ApJS..171..520C}. The ZEPHYR code computes the Alfv\'enic wave energy using a damped wave action conservation equation that contains the assumption that $Z_{-} \gg Z_{+}$. Thus, when reporting the magnitudes of the Els\"asser variables here for direct comparison with the BRAID results, we make use of Equation (56) of \citet{2007ApJS..171..520C} and correct for non-WKB effects by multiplying these quantities by a factor of $(1 + {\cal R}^{2})/(1 - {\cal R}^{2})$, where ${\cal R}$ is the reflection coefficient. With this correction, there is good agreement between the modeling of the PCH for both codes. We use the same correcting factor when plotting the $\Delta{v_{\rm rms}}$ in Figure \ref{fig:elsa}b. Because ZEPHYR assumes equipartition, $\Delta{v_{\rm rms}}$ and $\Delta{b_{\rm rms}}$ are equivalent for that model.

\begin{figure}[!ht]
\includegraphics[width=\columnwidth]{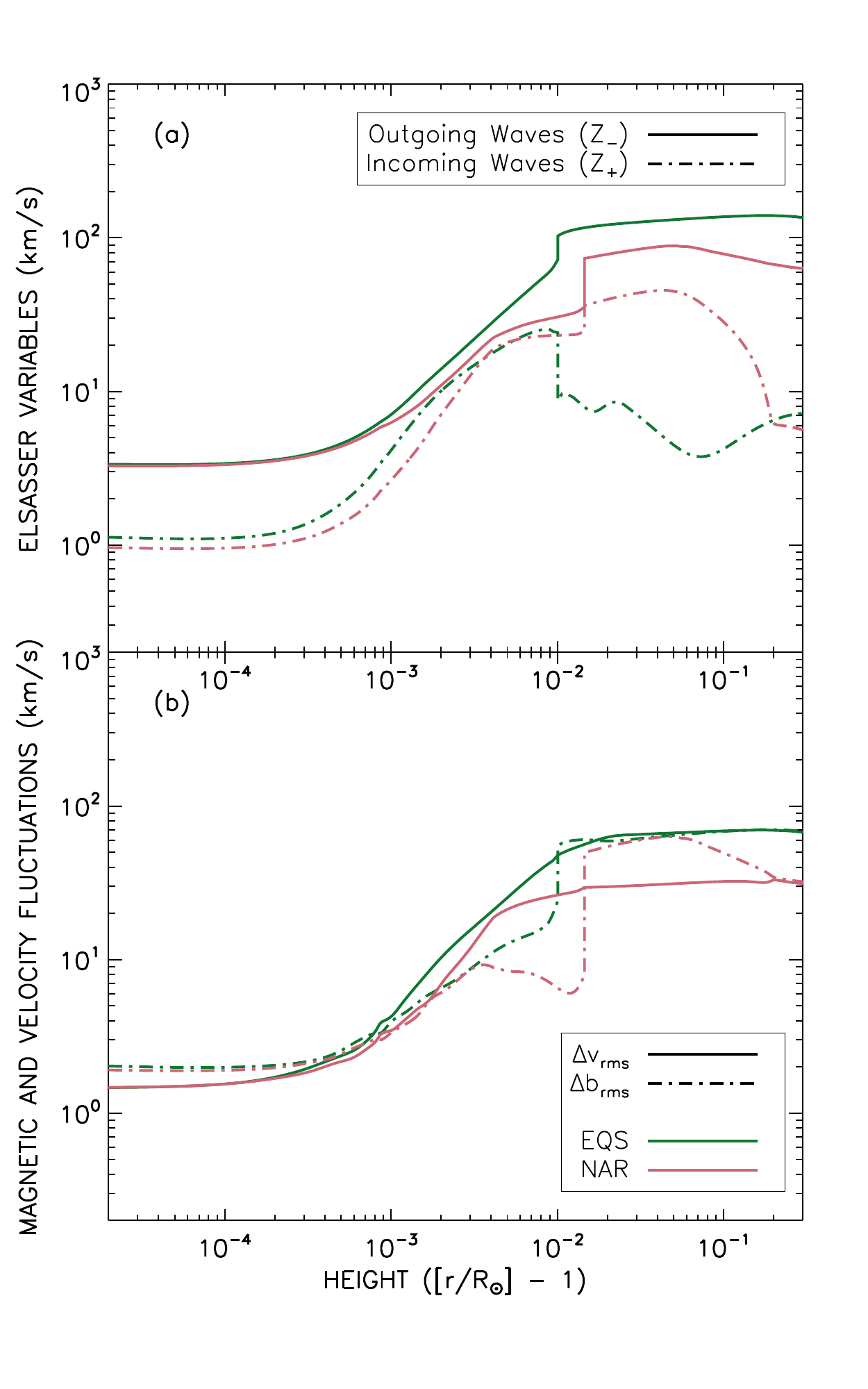}
\caption{Similar to Figure \ref{fig:elsa}; For the EQS and NAR models, we show the (a) outgoing and incoming Els\"asser variables and (b) the rms amplitudes of magnetic and velocity fluctuations.}
\label{fig:elsasser}
\end{figure}

Figure \ref{fig:elsasser} shows the Els\"asser variables and the amplitudes of magnetic and velocity fluctuations for both the equatorial streamer (EQS) and active region (NAR) models to a height of $z = 0.5 R_{\odot}$, where turbulence has had time to develop in the simulation. It is worthy of note that the NAR model shows an increase in $Z_{+}$ above the transition region where the PCH and EQS models show a decrease. For these models in Figure \ref{fig:elsasser}b and the results for the PCH (Figure \ref{fig:elsa}b), the shape of the $\Delta{b_{\rm rms}}$ is expected to have a sharp increase at the transition region, while the $\Delta{v_{rms}}$ doesn't show it \citep[see, e.g., Figure 9 of][]{2005ApJS..156..265C}. 

\subsection{Energy partitioning in the corona}

\begin{figure}[!ht]
\includegraphics[width=\columnwidth]{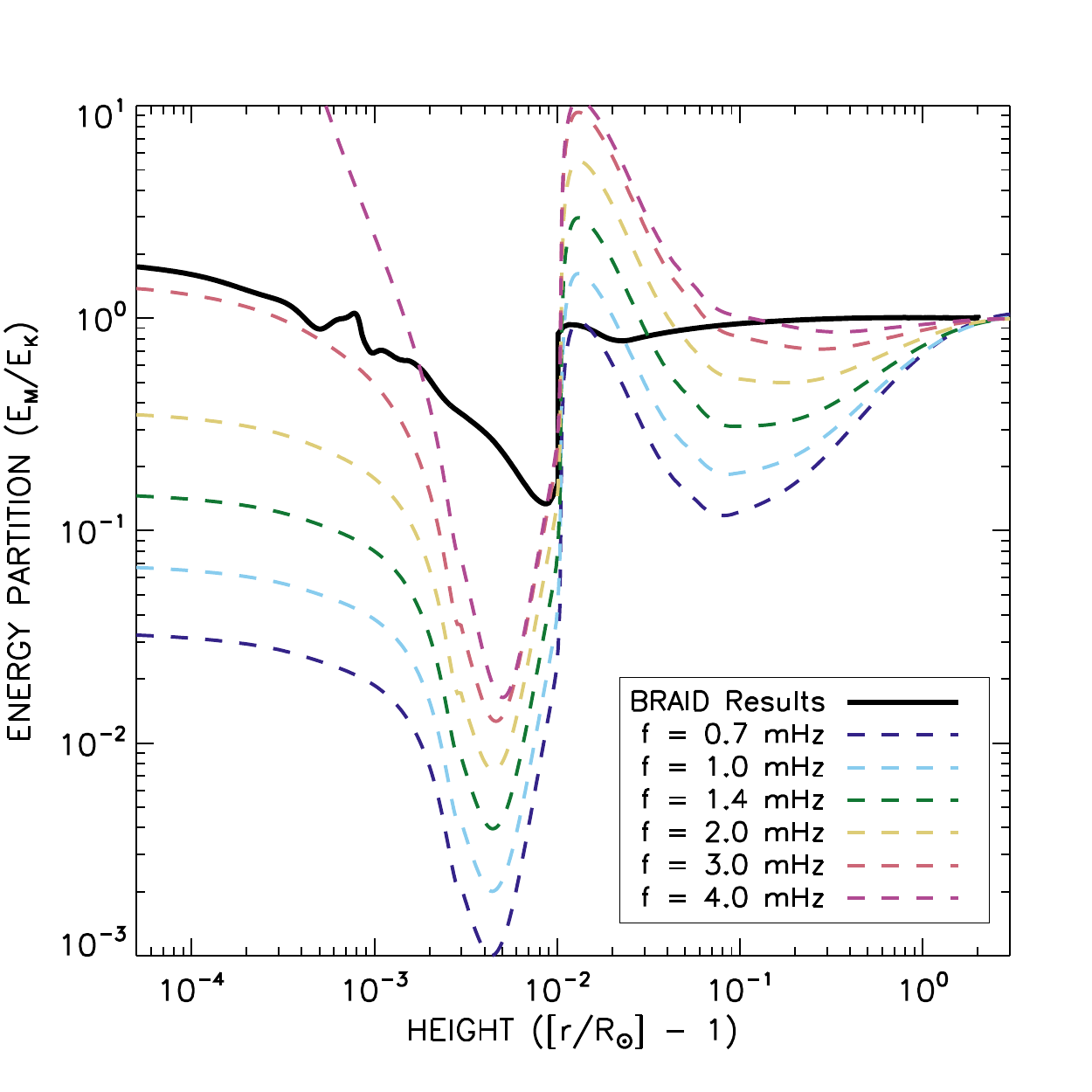}
\caption{Ratio of magnetic to kinetic energy density differs from equipartition below the transition region.}
\label{fig:part}
\end{figure}

In our previous modeling using ZEPHYR (see Section 2.1), we assumed equipartition between kinetic and magnetic potential energy densities. With BRAID, we are able to investigate how far the PCH model differs from this simplifying assumption. Figure \ref{fig:part} shows the ratio of the time-averaged magnetic and kinetic energy densities. Above the transition region, equipartition is a valid assumption. However, at the base of the flux tube, magnetic potential energy dominates, and this transitions to a stronger dominance of the kinetic energy right up to the transition region.

Also plotted in Figure \ref{fig:part} are six curves showing predictions from non-WKB reflection with a range of frequencies between 0.7 and 4.0 mHz. While the general shape is consistent with the BRAID results, it is interesting to note that the linear method of computing non-WKB reflection does have its limits. These predictions were made using the coronal hole model from \citet{2005ApJS..156..265C} as a basis. However, that model had a lower transition region ($z_{TR} \approx 0.003 R_{\odot}$), so we have multiplied the height coordinate by a factor of 3.4 to match with the ZEPHYR/BRAID coronal hole model used in this project.

Non-WKB theory predicts the kinetic energy to dominate in the chromosphere (e.g., $E_{M}/E_{K} < 1$ at low heights). This was discussed in Appendix A of \citet{2005ApJS..156..265C} as an after-effect of the transition from kink-mode MHD waves in the photosphere to volume-filling Alfv\'{e}n waves in the upper chromosphere. At low frequencies, the kink-mode waves are partially evanescent, with two possible solutions for the height-increase of $\Delta v_{\rm rms}$. One of the solutions has $E_{M}/E_{K} < 1$ and the other has $E_{M}/E_{K} > 1$. However, only the solution with $E_{M}/E_{K} < 1$ has a physically realistic energy density profile (exponentially decaying with increasing height), and this solution also corresponds to a net upward phase speed \citep[see also][]{1995ApJ...444..879W}.

\subsection{Heating rates: comparison with time-steady modeling}
\begin{figure}[!ht]
\includegraphics[width=\columnwidth]{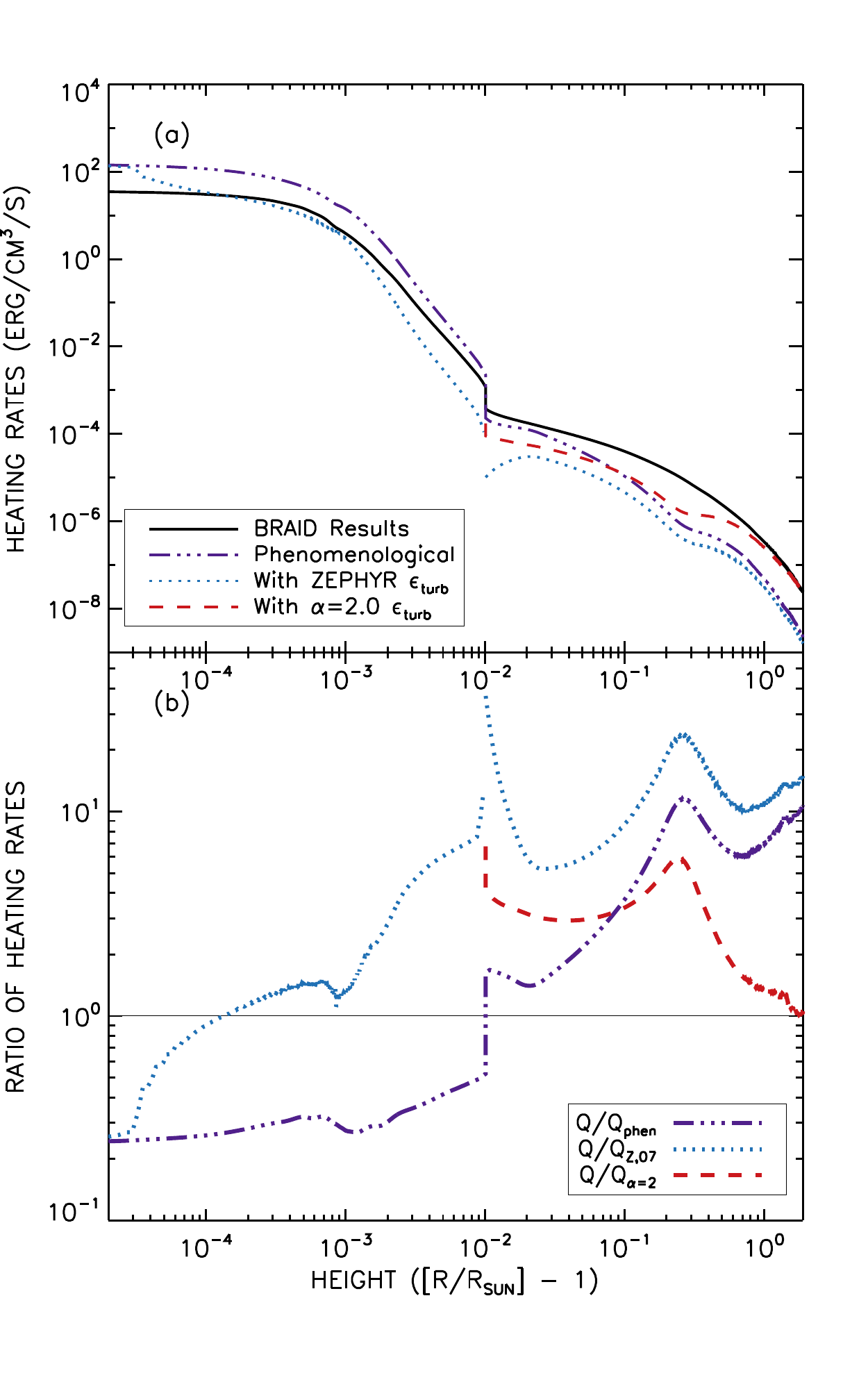}
\caption{We compare the time-averaged heating rate from BRAID for the PCH model with the phenomenological expressions given by Equations \eqref{eq:phen}, \eqref{eq:qfull} and \eqref{eq:looprate} by (a) directly plotting them as a function of height and (b) plotting ratios of the rates, with a thin black line showing unity.}
\label{fig:Qcompare}
\end{figure}

We compare the time-averaged heating rates from BRAID with the results from the time-steady modeling using ZEPHYR. Figure \ref{fig:Qcompare}a shows the radial dependence of the heating rate $Q$, and Figure \ref{fig:Qcompare}b shows the ratio between the numerically computed heating rates $Q$ with the phenomenological heating rate $Q_{\rm phen}$, as well as comparisons between $Q$ and phenomenological heating rates with added correction factors ($Q_{\rm Z,07}$ and $Q_{\rm loop}$), described in the following paragraphs. The heating rate $Q_{\rm phen}$ is based on the result of a long series of turbulence simulations and models \citep{1980PhRvL..45..144D,1983A&A...126...51G,1995PhFl....7.2886H,1999ApJ...523L..93M,2001ApJ...548..482D,2002ApJ...575..571D,2009ApJ...701..652C}. The analytical expression for this base phenomenological rate is given by: \begin{equation}Q_{\rm phen} = \rho_{0}\frac{Z_{-}^{2}Z_{+} + Z_{+}^{2}Z_{-}}{4L_{\perp}}\label{eq:phen}\end{equation}
where $\rho_{0}$ is the solar wind density, $Z_{-}$ and $Z_{+}$ are the Els\"asser variable amplitudes that represent incoming and outgoing Alfv\'en waves, and $L_{\perp}$ is the turbulent correlation length. We normalize $L_{\perp}$ to a value of 75~km at the photosphere \citep{2007ApJS..171..520C}. This allows us to write $L_{\perp} = 0.75 R$, where $R$ is the radius of the flux tube, which scales as $B^{-1/2}$. Below the transition region, $Q/Q_{\rm phen} \approx 0.2$, which is similar to what \citet{2011ApJ...736....3V} found for closed loops. Above the transition region, $Q/Q_{\rm phen} \gg 1$, which may be explainable if the actual correlation length $L_{\perp}$ expands {\it less rapidly} than we assumed from its proportionality with the flux tube radius $R$. Because the ratio $Q/Q_{\rm phen}$ strays as much as an order of magnitude away from unity, we further compare BRAID's computed heating rate with analytical expressions that contain correcting factors that take into account efficiency of turbulence as a function of height.

The first of the two efficiency factors we use is based on the prior work of \citet{2001ApJ...548..482D}, \citet{2003ApJ...597.1097D}, and \citet{2007ApJS..171..520C}. The extended expression, $Q_{\rm Z,07}$ is given by: \begin{subequations}
\begin{align}
Q_{\rm Z,07} = \epsilon_{\rm turb} Q_{\rm phen},\\
\epsilon_{\rm turb} = \frac{1}{1+(t_{\rm eddy}/t_{\rm ref})^{n}},
\end{align}\label{eq:qfull}
\end{subequations} where $t_{\rm eddy}$ is the outer-scale eddy cascade time, $t_{\rm eddy} = L_{\perp}\sqrt{3\pi}/(1+M_{\rm A})/v_{\perp}$, and $t_{\rm ref}$ is the macroscopic Alfv\'en wave reflection timescale, $t_{\rm ref} = 1/|\nabla \cdot {\bf V_{\rm A}}|$ \citep{2007ApJS..171..520C}. In the expression for $t_{\rm eddy}$, the velocity $v_{\perp}$ is the amplitude of perpendicular fluctuations, defined previously in the BRAID results as $\Delta v_{\rm rms}$. When $t_{\rm eddy} \gg t_{\rm ref}$, turbulent heating is quenched. The turbulent efficiency factor $\epsilon_{\rm turb}$ accounts for regions where energy is carried away before a turbulent cascade can develop. The exponent $n$ is set to 1 based on analytical and numerical models by \cite{1980PhRvL..45..144D}, \cite{1989PhFlB...1.1929M}, and \cite{2006PhPl...13d2306O}. The efficiency factor works to make $Q_{\rm Z,07} < Q_{\rm phen}$, bringing the ratio $Q/Q_{\rm Z,07}$ up relative to $Q/Q_{\rm phen}$. At low heights, where the efficiency factor is low, the inclusion of the efficiency factor defined in Equation \eqref{eq:qfull}b does help to bring $Q$ and $Q_{\rm Z,07}$ into better agreement. At large heights, the efficiency factor is closer to 1, so $Q_{\rm phen}$ and $Q_{\rm Z,07}$ {\it both} underestimate the heating rate computed by BRAID.

An alternative efficiency factor has emerged from studies of closed coronal loops driven by slow transverse footpoint motions. In such models, magnetic energy is built up from the twisting and shearing motions of the field lines \citep{1972ApJ...174..499P}, and the energy dissipation appears to follow a cascade-like sequence of quasi-steady relaxation events. \citet{2009ApJ...706..824C} parametrized the time-averaged heating rate in these
models as \begin{equation}
  Q_{\rm loop} = \epsilon_{\rm loop} Q_{\rm phen}
  \,\,\, , \,\,\,\,\,\,\,\,\,
  \epsilon_{\rm loop} \, = \,
  \left( \frac{L_{\perp} V_{\rm A}}{v_{\perp} L_{\parallel}}
  \right)^{\alpha}\label{eq:looprate}
\end{equation} where $L_{\parallel}$ is often defined as the ``loop length'' for closed magnetic structures, and $\alpha$ is an exponent that describes the sub-diffusive nature of the cascade in a line-tied loop. The quantity in parentheses is a ratio of timescales; this ratio is the nonlinear time over the wave travel time. To set the value of $L_{\parallel}$ in our open-field models, we follow \citet{2004ApJ...615..512S}, who found that open and closed regions can be modeled using a unified empirical heating parametrization when the actual loop length $L$ is replaced by an effective length scale \begin{equation}  L_{\parallel} \, \approx \, \frac{L}{1 + (L/L_{0})} \end{equation} where $L_{0} = 50$~Mm.Thus, for open-field regions in which $L \gg L_0$, we use $L_{\parallel} = L_0$.

The exponent $\alpha$ describes how interactions between
counter-propagating Alfv\'{e}n wave packets can become modified by MHD processes such as scale-dependent dynamic alignment \citep{2009ApJ...699L..39B}. The value $\alpha = 0$ corresponds to a classical hydrodynamic cascade. \citet{2000SoPh..195..299G} constructed a model of MHD turbulence in which $\alpha = 1.5$, and \citet{1986ApJ...311.1001V} constructed a random-walk type model in which $\alpha = 2$ \citep[see also][]{2008ApJ...682..644V}. \citet{2008ApJ...677.1348R}, however, found from numerical simulations that $\alpha$ can occupy any value between 1.5 and 2, depending on the properties of the background corona and wave driving. \citet{2009ApJ...706..824C} created an analytic prescription for specifying $\alpha$ in a way that agrees with the \citet{2008ApJ...677.1348R} results. Subsequently, \citet{2009ApJ...706..824C} and \citet{2013ApJ...772..149C} found that when modeling the coronal X-ray emission from low-mass stars, the longest loops---which seem to be most appropriate to compare with open-field regions---tend to approach the high end of the allowed range of exponents (i.e., $\alpha \approx 2$). Thus, in this paper we use $\alpha = 2$ and note that the differences between $Q_{\rm loop}$ and $Q_{\rm phen}$ will be reduced for smaller values of the exponent. At the largest heights, $Q_{\rm loop}$ better matches the BRAID numerical results than either $Q_{\rm phen}$ or $Q_{\rm Z,07}$.

\begin{figure}[!ht]
\includegraphics[width=\columnwidth]{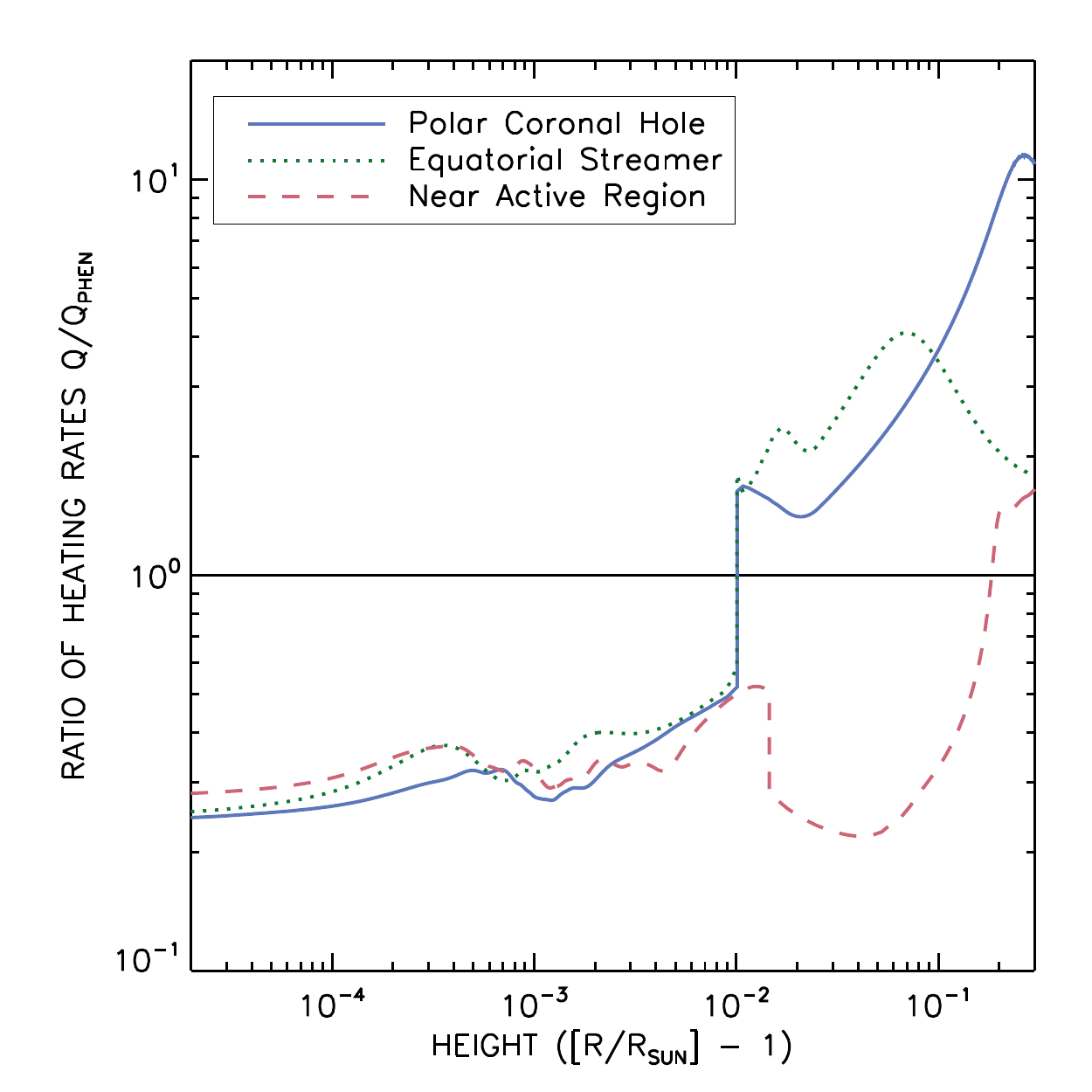}
\caption{For each model, we show the ratio of the time-averaged heating rate from BRAID with the phenomenological expression given by Equations \eqref{eq:phen}.}
\label{fig:Qother}
\end{figure}

For additional comparison, in Figure \ref{fig:Qother} we plot the ratio of BRAID numerical results with the phenomenological heating rate for all three of our flux tube models. The behavior of the heating rate expressions with efficiency factors is similar, so we show only the ratio $Q/Q_{\rm phen}$ in Figure \ref{fig:Qother}. Note that the EQS model behaves similarly to the PCH model, but the NAR model exhibits a marked decrease in the ratio $Q/Q_{\rm phen}$ in the low corona before increasing again to approach the EQS model at the top of the grid. This behavior is reminiscent of the closed-field models of \citet{2011ApJ...736....3V}, in which $Q/Q_{\rm phen}$ came back down to values of 0.2--0.3 in the coronal part of the modeled loops.

\section{Extended analysis of time variability}

\subsection{Statistical variations as a source of multithermal plasma}

\begin{figure}[!ht]
\includegraphics[width=\columnwidth]{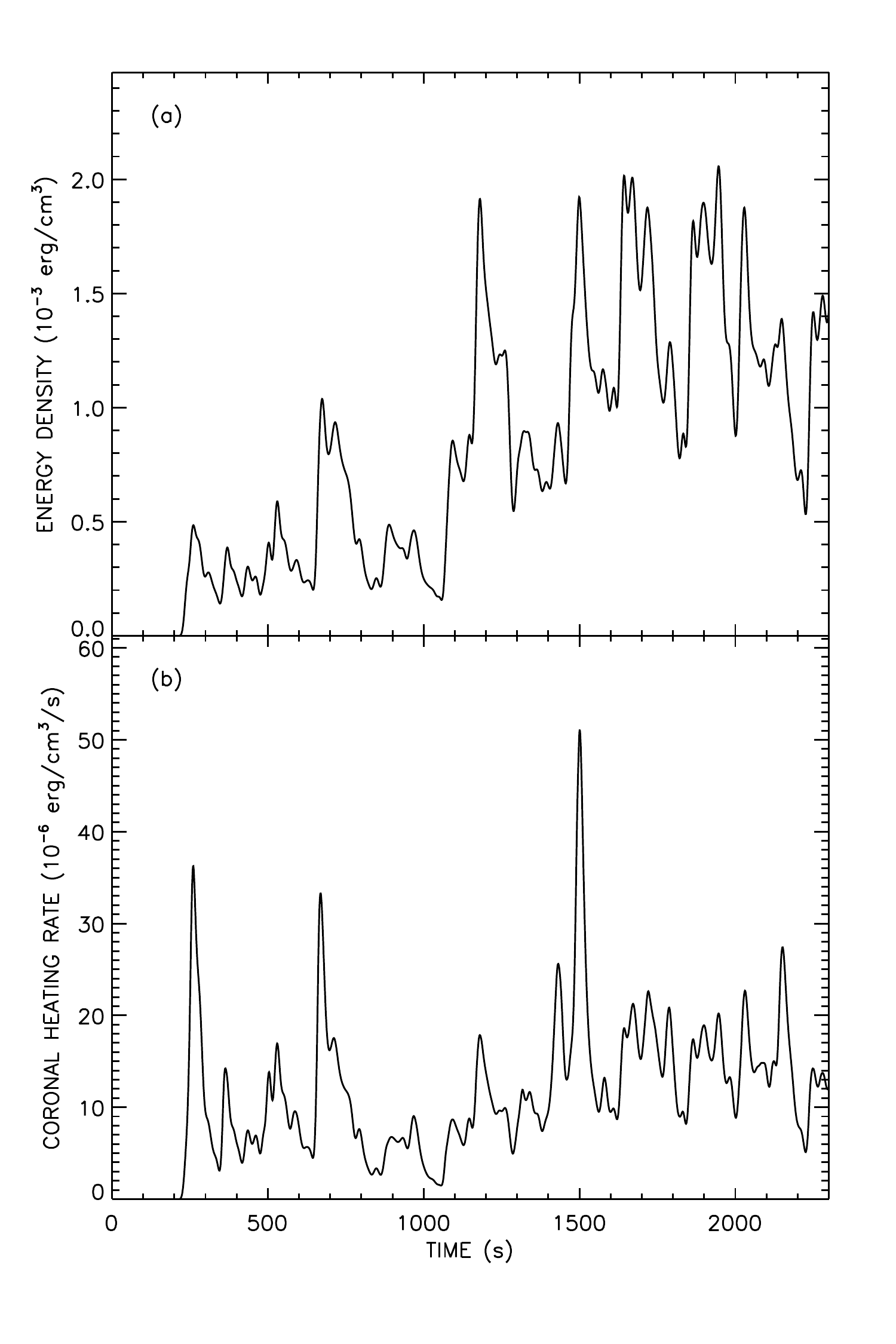}
\caption{Due to the random jostling that generates the Alfv\'en waves, the (a) spatially-averaged energy density in the corona and (b) spatially-averaged heating rate per unit volume in the corona are bursty and strongly time-dependent in nature.}
\label{fig:bursty}
\end{figure}

The turbulent heating simulated by BRAID was found by \citet{2011ApJ...736....3V} to be quite intermittent and variable on small scales. Figure \ref{fig:bursty} illustrates some of this variability by showing the fluctuation energy density and heating rate volume-averaged over the low corona (i.e., between the transition region at $z = 0.01 \, R_{\odot}$ and an upper height of $0.5 \, R_{\odot}$). This is a similar plot as Figure 4 of \citet{2011ApJ...736....3V}.  Even with this substantial degree of spatial averaging, the nanoflare-like burstiness generated by the turbulence is evident in Figure \ref{fig:bursty}. There is a large body of prior work concerning such intermittent aspects of turbulent heating \citep[see, e.g.,][]{1996ApJ...457L.113E,1996ApJ...467..887H,1997ApJ...484L..83D,1998ApJ...505..974D,1999PhPl....6.4146E}.

The time-varying heating rate should also give rise to a similarly variable coronal temperature structure. We investigate the possibility that the resulting stochastic distribution of temperatures may be partially responsible for the observational signatures of {\em multithermal} plasmas---e.g., nonzero widths of the differential emission measure (DEM) distribution. \citet{2012ApJ...746...81A} studied the spatial and temporal response of a conduction-dominated corona to the simulated variations in $Q$ from BRAID. They found that conduction leads to a ``smeared out'' temperature structure that nevertheless retains much of the bursty variability seen in the heating rate. Here, we perform an even simpler estimate of the distribution of temperatures by taking the distribution of volume-averaged heating rates $\langle Q \rangle$ shown in Figure \ref{fig:bursty} and processing each value through the simple conductive scaling relation of \citet{1978ApJ...220..643R}.  Thus, \begin{equation}
  \frac{\langle T \rangle}{\langle \bar{T} \rangle} \, = \, \left(
  \frac{\langle Q \rangle}{\langle \bar{Q} \rangle} \right)^{2/7}
\end{equation} where $\langle T \rangle$ is an estimated volume-averaged coronal temperature. The normalizing value of the heating rate $\langle \bar{Q} \rangle$ is assumed to be the mean value of $\langle Q \rangle$ seen in the BRAID simulation. For simplicity, we take the normalizing value of the temperature $\langle \bar{T} \rangle$ to be the maximum coronal temperature found in the corresponding ZEPHYR model from \citet{2007ApJS..171..520C}. The PCH, EQS, and NAR models exhibited values of $\langle \bar{T} \rangle$ of 1.352, 1.224, and 1.675 MK, respectively.

\begin{figure}[!ht]
\includegraphics[width=\columnwidth]{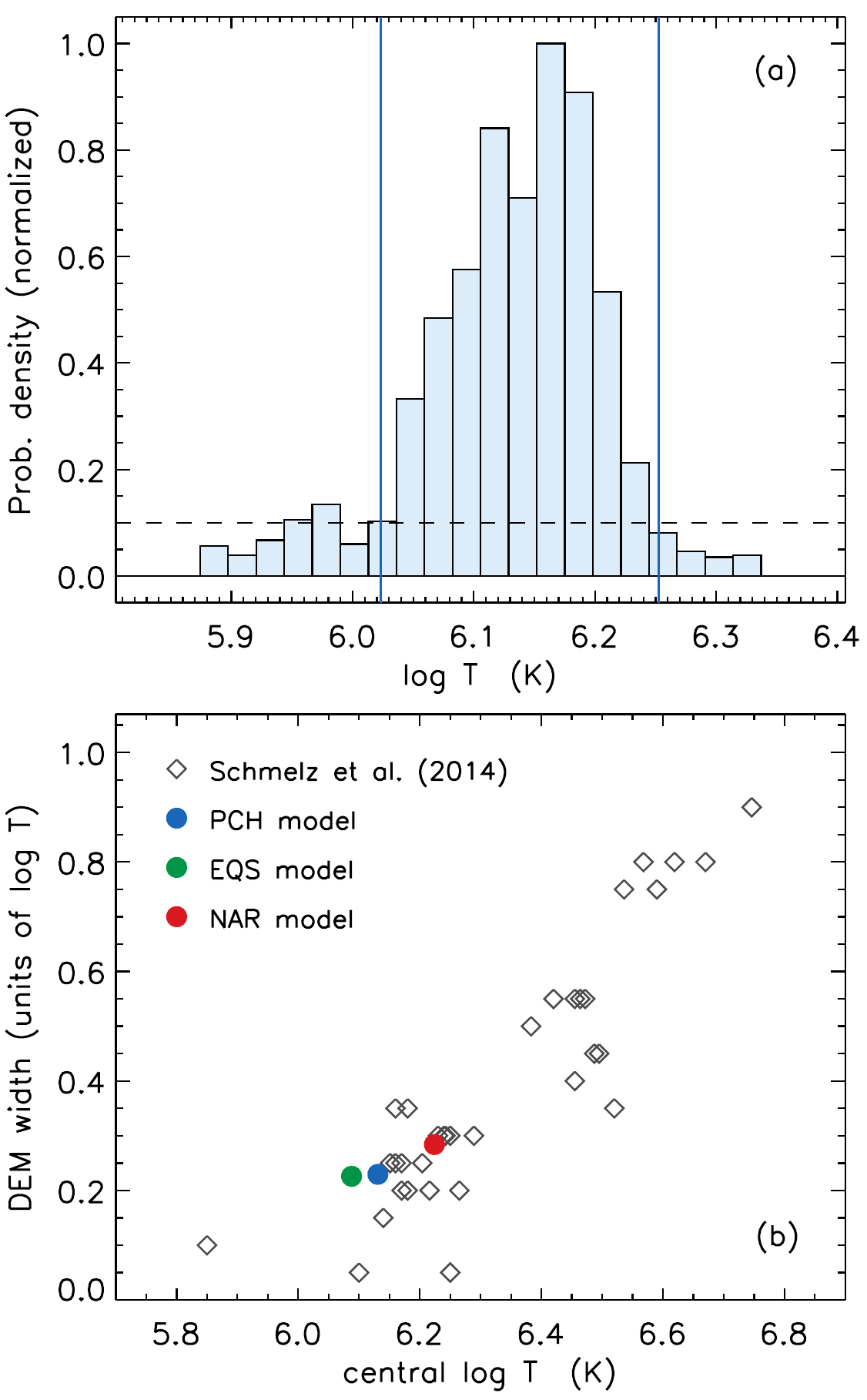}
\caption{(a) Scaled histogram of the probability distribution of coronal temperatures estimated from BRAID heating rates. Temperatures at which the distribution drops to 0.1 times the maximum value (dashed line) are indicated by solid red lines.
(b) Observed DEM widths from \citet{2014ApJ...795..171S}, plotted as the relative width in $\log T$ space and computed from the reported number of $\log T$ ``bins'' (gray diamonds), compared with simulated DEM widths from the PCH (red circle), EQS (green circle), and NAR (blue circle) models.}
\label{fig:dem}
\end{figure}

Figure \ref{fig:dem}(a) shows the distribution of derived values of $\langle T \rangle$ for the PCH model. If the temperatures along this flux tube were measured by standard ultraviolet and X-ray diagnostics, with time integrations long in comparison to the scale of variability in the BRAID model, then this distribution would be equivalent to the DEM. For each simulated DEM, we measured its representative ``width'' in the same way as described by \citet{2014ApJ...795..171S}; i.e., we used the points at which the DEM declined to 0.1 times its maximum value. For the PCH, EQS, and NAR models, we found widths of 0.2294, 0.2259, and 02839 in units of ``dex'' ($\log T$), respectively.

Figure \ref{fig:dem}(b) compares the properties of the three simulated DEMs with a selection of observationally derived coronal-loop DEMs from \citet{2014ApJ...795..171S}. The BRAID models do appear to reproduce the observed multithermal nature of coronal plasmas, both in the absolute values of the widths (which fall comfortably within the range of the observed values) and in the overall trend for hotter models to have broader DEMs. Of course, the open-field models studied here only span a very limited range of central temperatures $\langle \bar{T} \rangle$ in comparison to the observed cases.

\subsection{Power spectrum of fluctuations}

For the PCH model, we investigated in detail the power spectrum of the velocity fluctuations caused by the Alfv\'en waves. To generate the Fourier transform, we assumed constant time spacing using the model results spanning from $t = \tau_{\rm A}$ to $t = 2 \tau_{\rm A}$. We subtracted the mean, doubled the length to make a periodic sequence, and then fed the cleaned quantity into a traditional FFT procedure. The power spectrum is the product of the result of that FFT procedure with its complex conjugate. This procedure gives us these spectra at each height $z$.

\begin{figure}[!ht]
\includegraphics[width=\columnwidth]{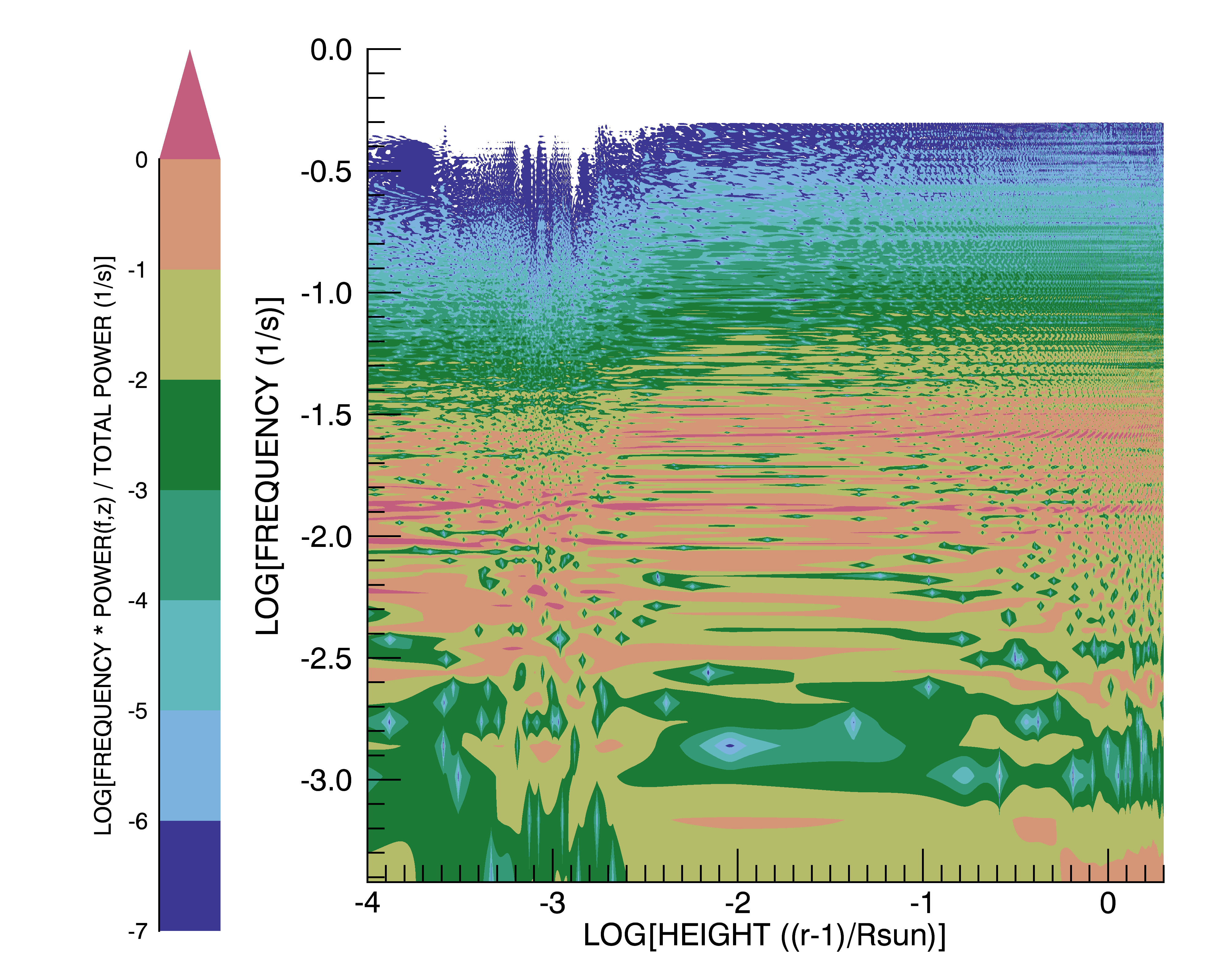}
\caption{Contour plot of the power spectrum in the PCH model showing power in color, as a function of height and of frequency.}
\label{fig:contour}
\end{figure}

\begin{figure}[!ht]
\includegraphics[width=\columnwidth]{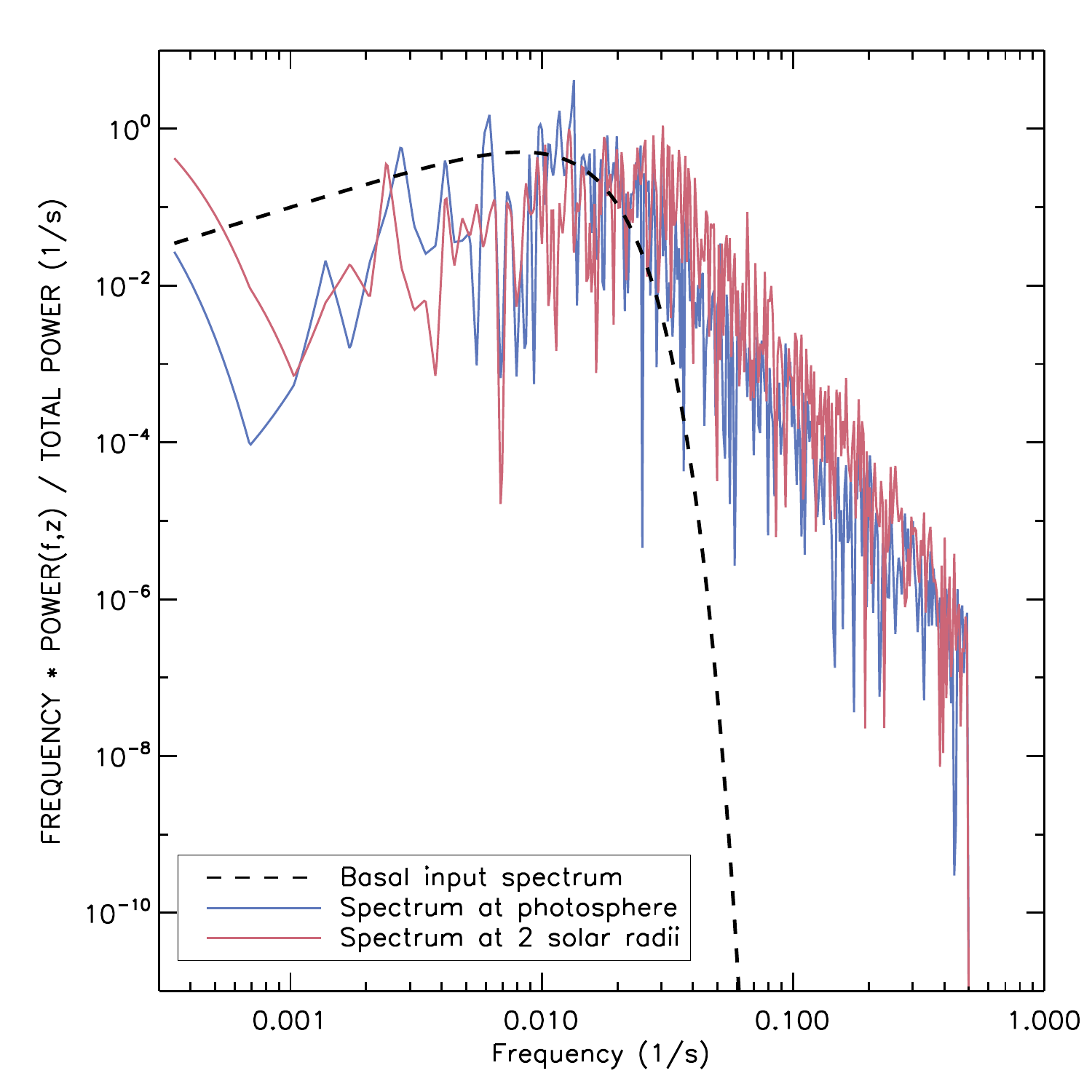}
\caption{Power spectra for the upper and lower boundaries of the PCH model compared to the input spectrum show a boost in power at high frequencies.}
\label{fig:spectra}
\end{figure}

In Figure \ref{fig:contour}, we examine a contour plot of the power in these fluctuations. Certain frequencies in the $10^{-2}$ Hz to $10^{-1.5}$ Hz range have a relatively high amount of power at all heights, while at the higher frequencies, there is an increase in power as a function of height. This is more easily seen in Figure \ref{fig:spectra}. We show the basal input spectrum with a dashed line, and the power spectrum at the upper boundary of our model lies above the power spectrum at the photospheric base for the highest frequencies. Both boundaries show that there is a boost above the input spectrum for high frequencies.

Figure \ref{fig:spectra} shows that the high-frequency part of the BRAID turbulent power spectrum appears to be a power law ${\cal P} \propto f^{-n}$, where the value of $n$ varies a bit with height in the model. Looking only at the FFT data with $f \geq 0.03$~Hz, we found that $n \approx 4.5$ at the photospheric base, and then it steepens at larger heights to take on values of order 5.2--6.4 (i.e., a mean value of 5.8 with a standard deviation of $\pm$0.6) at chromospheric heights below the TR. In the corona, however, $n$ decreases a bit to a mean value of 4.9 and a smaller standard deviation of $\pm$0.3. Some of the quoted standard deviations are likely due to fitting uncertainties of the inherently noisy power spectra, but it is clear that the corona exhibits less variation in $n$ than the regions below the TR.

Despite several decades of spacecraft observations of power-law frequency spectra in the turbulent solar wind, it was not immediately obvious that the BRAID power spectrum should have exhibited such power-law behavior. Spacecraft-frame measurements are often interpreted as being a spatial sample through quasi-stationary wavenumber variations \citep[see, e.g.,][]{1938RSPSA.164..476T,2012SSRv..172..325H}. In strong-field MHD turbulence, the dominant wavenumber cascade is expected to be in the $k_{\perp}$ direction. However, the Alfv\'{e}nic fluctuations that make up an MHD cascade have a dispersion relation in which the frequency depends primarily on $k_{\parallel}$. Thus, MHD (especially RMHD) turbulence is typically described as ``low-frequency turbulence'' and the idea of an inherent power-law {\em frequency cascade} is met---usually, rightly so---with skepticism.

There has been one proposed model in which a power-law spectrum in frequency (i.e., in $k_{\parallel}$) occurs naturally and without the need for substantial parallel cascade: the so-called ``critical balance'' model of \citet{1995ApJ...438..763G}. In this picture, strong mixing is proposed to occur between the turbulent eddies (primarily moving perpendicular to the background field) and Alfv\'{e}n wave packets (moving parallel to the field), such that the parameter space ``filled'' by a fully developed cascade
is determined by the critical balance parameter \begin{equation}\chi \, = \, \frac{k_{\parallel} V_{\rm A}}{k_{\perp} v_{\perp}}\end{equation} taking on values $\chi \lesssim 1$ \citep[see also][]{1984ApJ...285..109H}. The limit $\chi \ll 1$ corresponds generally to low-frequency fluctuations with $k_{\parallel} \ll k_{\perp}$ as is expected in anisotropic MHD turbulence. This parameter can also be interpreted as a ratio of timescales. 

Following the implications of critical balance led \citet{1995ApJ...438..763G} to a phenomenological expression of the time-steady inertial range, given here as \begin{equation} E (k_{\parallel},k_{\perp}) \, = \, \frac{V_{\rm A} \, v_{\perp}(k_{\perp})}{k_{\perp}^3} \, g (\chi) \label{eq:E3D}\end{equation} where this combines Equations (5) and (7) of \citet{1995ApJ...438..763G}. Above, $E$ is a three-dimensional power spectrum that gives the magnetic energy density variance (in velocity-squared units) when integrated over the full volume of wavenumber space, \begin{equation} \frac{\langle B_{\perp}^{2} \rangle}{8\pi\rho} \, = \, \int d^{3} {\bf k} \,\, E (k_{\parallel},k_{\perp}) \, = \, \int df \,\, {\cal P}(f) \,\, ,\label{eq:eq11}\end{equation} and we also define the frequency spectrum ${\cal P}(f)$ in a similar way. Note that $2\pi f = \omega = k_{\parallel} V_{\rm A}$ in the local rest frame of the plasma under the assumption that the fluctuations are Alfv\'en waves.

In Equation (\ref{eq:E3D}) above, $v_{\perp} (k_{\perp})$ is the reduced velocity spectrum, which specifies the magnitude of the velocity
perturbation at length scales $k_{\perp}^{-1}$ and $k_{\parallel}^{-1}$. This spectrum is often assumed to be a power-law with $v_{\perp} \propto k_{\perp}^{-m}$. The exponent $m$ has been proposed to range between values of 1/3 \citep[strong turbulence;][]{1995ApJ...438..763G} and 1/2 \citep[weak turbulence;][]{2000JPlPh..63..447G}. Lastly, the function $g(\chi)$ in Equation (\ref{eq:E3D}) is a ``parallel decay'' function that is expected to become negligibly small for $\chi \gg 1$. Because $g(\chi)$ is normalized to unity when integrated over all $\chi$, a simple approximation for it is a step function,
\begin{equation}
  g(\chi) \, \approx \, \left\{
  \begin{array}{ll}
  1 \,\, , & \chi \leq 1 \,\, , \\
  0 \,\, , & \chi > 1
  \end{array}
  \right. 
\end{equation} \citep[see also][]{2002ApJ...564..291C,2012ApJ...754...92C}. The above form for $g(\chi)$, combined with the assumption that the reduced spectrum $v_{\perp}(k_{\perp})$ extends out to $k_{\perp} \rightarrow \infty$, leads to the high-frequency end of the frequency spectrum obeying a power law, with \begin{equation} {\cal P}(f) \, \propto \, f^{(m+1)/(m-1)} \,\, .\end{equation} The strong turbulence case ($m=1/3$, ${\cal P} \propto f^{-2}$) has been studied extensively both observationally and theoretically \citep[see, e.g.,][]{2012SSRv..172..325H}.

Of course, the BRAID models highlighted in this paper do {\em not} have reduced perpendicular velocity spectra that extend over large ranges of $k_{\perp}$ space. The models presented here (similar to those of \citet{2011ApJ...736....3V}) resolve only about one order of magnitude worth of an ``inertial range'' in $k_{\perp}$ space. For the step-function version of $g(\chi)$ given above, the imposition of a $v_{\perp}(k_{\perp})$ cutoff above an arbitrary $k_{\rm max}$ produces a frequency spectrum ${\cal P}(f)$ that is similarly cut off above a frequency $f_{\rm max}$ determined by critical balance ($\chi = 1$) at $k_{\perp} = k_{\rm max}$.

In an alternative to the step-function version of $g(\chi)$, \citet{2003ApJ...594..573C} found an analytic solution for $g(\chi)$ by applying an anisotropic cascade model that obeyed a specific kind of advection--diffusion equation in three-dimensional wavenumber space. The general form of $g(\chi)$ is reminiscent of a suprathermal kappa function \citep[see, e.g.,][]{2010SoPh..267..153P}, which is roughly Gaussian at low $\chi$ and a power-law at large $\chi$. For $\chi \gg 1$, \citet{2003ApJ...594..573C} found that $g(\chi) \propto \chi^{-(3s+4)/2}$, where $s$ is the ratio of the model's perpendicular advection coefficient to the perpendicular diffusion coefficient. It is still not known if MHD turbulence in the solar corona and solar wind exhibits a universal value of $s$, or even whether or not $s$ is even a physically meaningful parameter. Nevertheless, the wavenumber diffusion framework of \citet{1990JGR....9514881Z} and \citet{2009PhRvE..79c5401M} has been shown to be consistent with a value of $s=2$ in this family of advection--diffusion equations. In a different model of coronal turbulence, \citet{1986ApJ...311.1001V} showed that a cascade of slow random-walk displacements of the field lines can be treated as the case $s=1$. On the basis of observations alone, \citet{2003ApJ...594..573C} and \citet{2009ApJ...691..794L} found that if $s$ could be maintained at small values of order 0.1--0.3, there would be sufficient high-frequency wave energy to heat protons and minor ions via ion cyclotron resonance.

No matter the value of $s$ or the reduced spectral index $m$, it can be shown that at large frequencies ($f > f_{\rm max}$), a power law of the form $g(\chi) \propto \chi^{-n}$ produces a power-law frequency spectrum ${\cal P} \propto f^{-n}$ with the same exponent (see the integration over $k_{\perp}$ in Equation \eqref{eq:eq11}). Thus, we postulate that measuring $n$ from the BRAID simulations {\em may} be a way to extract information about the exponent $s$, with \begin{equation} s \, = \, \frac{2n - 4}{3} \,\, .\end{equation} The values of $n$ reported above imply a photospheric value of $s \approx 1.7$, which increases to $s \approx 2.5$ in the chromosphere (with a relatively large spread) and then decreases to $s \approx 1.9$ in the corona. The similarities to the theoretical value of $s=2$ \citep[from, e.g.,][]{1990JGR....9514881Z, 2009PhRvE..79c5401M} are suggestive, but not conclusive.

\subsection{Nanoflare statistics of heating rate variability}
\begin{figure}[h!]
\includegraphics[width=\columnwidth]{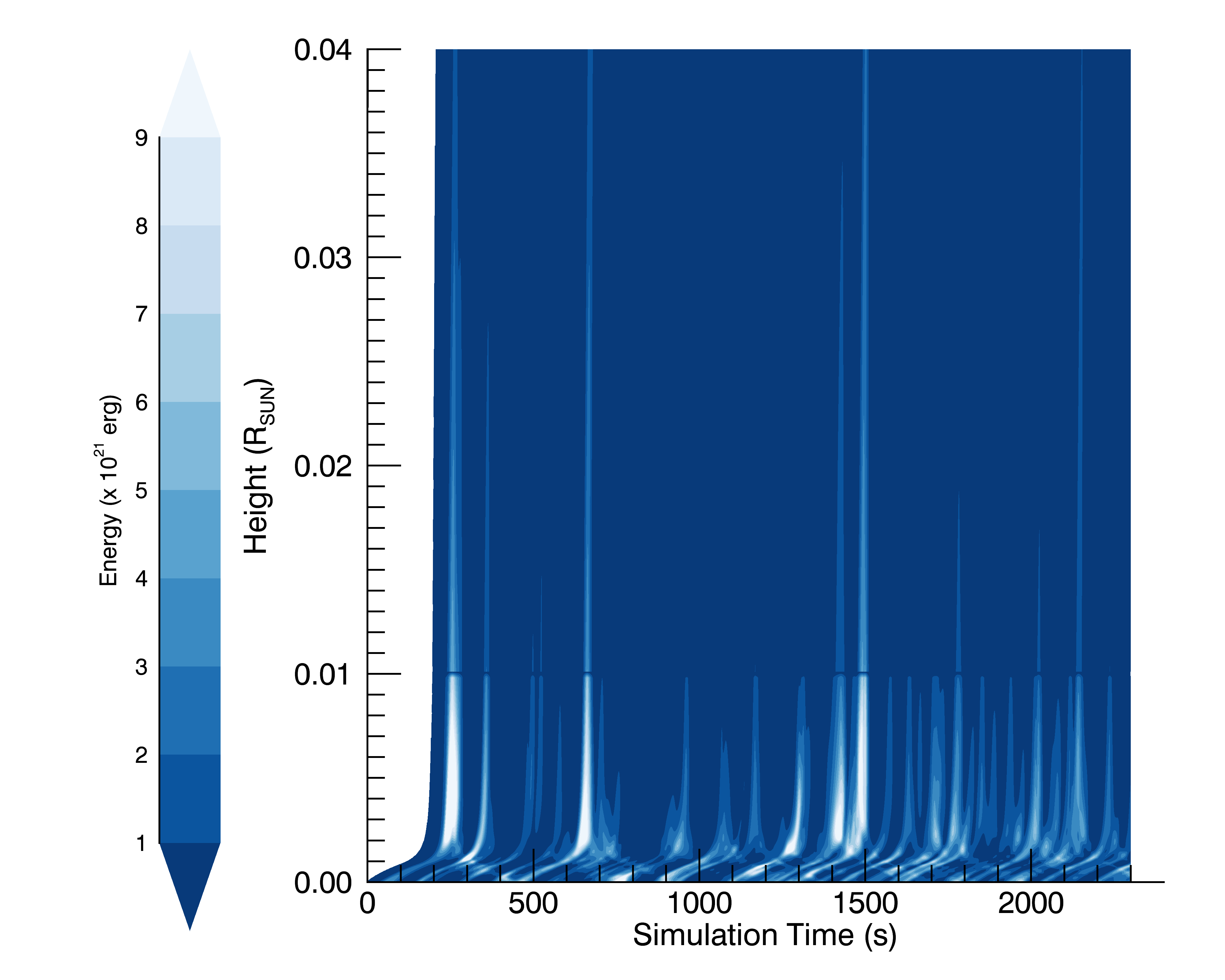}
\caption{Dependence of energy on height above photosphere (y-axis) and time (x-axis). The transition region is at 0.01 R$_{\odot}$.}
\label{fig:cenergy}
\end{figure}

We investigated the variability of heating and energy as a function of height and time throughout the simulation. In a given finite ``zone,'' the energy lost via dissipative heating can be calculated using the heating rate as
\begin{equation}E(z,t) = Q(z,t)\Delta z (\pi R^{2})\Delta t,\end{equation}
where the zones are defined at a set of heights, $z$, with unequal spacing $\Delta z$, and at a set of times, $t$, with equal spacing $\Delta t = 0.25$ s. In Figure \ref{fig:cenergy}, we provide a contour plot of the energy, comparable with Figure 6b of \citet{2013ApJ...773..111A}. Many of the impulsive heating events that result in spikes of energy over a short time frame stop at the transition region, which lies at $0.01 R_{\odot}$, but some extend to ten times that height. There is also a lack of energy at the beginning of the simulation ($t < \tau_{\rm A}$), where the information has not yet had time to propagate up through the model grid.

\begin{figure*}[!ht]
\includegraphics[width=\textwidth]{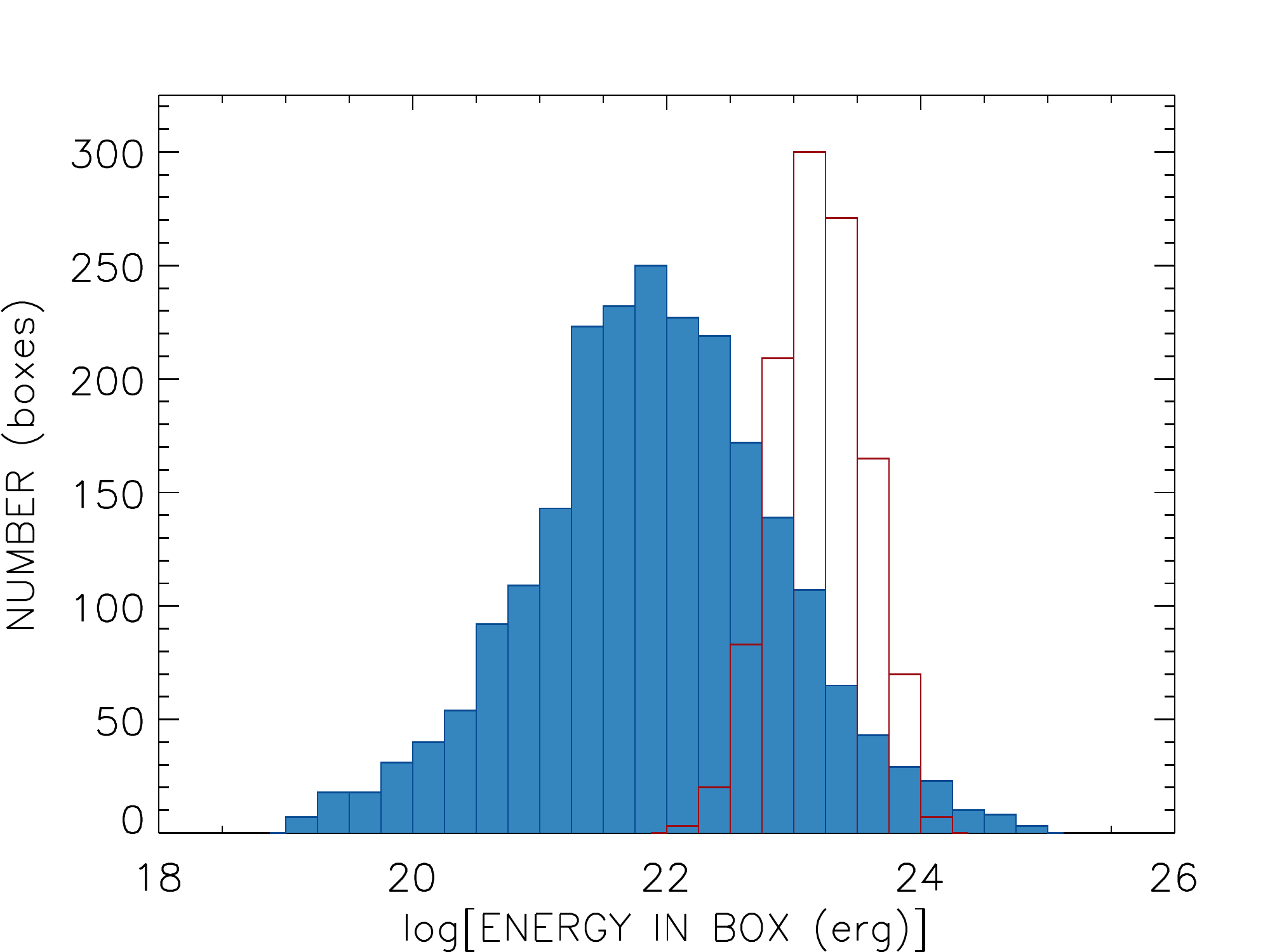}
\caption{Filled histogram shows the statistical distribution of energy contained in boxes of space and time for the PCH model. Only boxes above the transition region are used. For comparison, red outlined histogram shows the results from closed magnetic loops \citep{2013ApJ...773..111A}.}
\label{fig:denergy}
\end{figure*}

Following the method of \citet{2013ApJ...773..111A}, we then use box capturing to get a statistical sense of the distribution of energy in these zones throughout the corona. In order to be directly comparable to their defined events, we also use boxes with a width of 19.4 seconds in simulation time and height of 19.4 seconds in Alfv\'en travel time (a proxy for height). This choice in box size results initially in 118 sections across the time dimension and 39 sections along the height dimension. However, the lowest 10 boxes are at heights below the transition region, and we plot only boxes in the corona in Figure \ref{fig:denergy}. Additionally, we take out the first 770 seconds corresponding to one Alfv\'en travel time in the PCH model (recall Table 1 and Equation \eqref{eq:traveltime}) to ensure that the waves have had time to propagate fully throughout the corona.

These cuts result in 78 time sections and 29 height sections, giving us a total of 2262 coronal boxes. The arithmetic mean in log-space of the energy contained within these boxes is 21.91 with a standard deviation of 0.97. The average energy contained in the boxes is lower than that found by \citet{2013ApJ...773..111A}, since our model does not have two strong footpoints supplying separate sources of counter-propagating Alfv\'en waves, but it is still a significant amount of energy. Figure \ref{fig:denergy} also shows that we find a few events that reach higher energies than the histogram from \citet{2013ApJ...773..111A}, and that higher tail gets well into the classical nanoflare expected energies \citep{2011SSRv..159..263H}. The minimum-energy event contains 10$^{19.04}$ erg while the maximum-energy event contains 10$^{24.97}$ erg. It is worthy of note that the peak energies captured in these boxes fall well within the ``picoflare'' range \citep{1999SSRv...87...25A,2000ApJ...529..554P}. All of these events are a natural product of the total heating coming out of BRAID, suggesting that the physics contained within these models may lead to the formation of pico- and nanoflares.

\section{Discussion and Conclusions}

We have analyzed three typical open magnetic field structures using one-dimensional time-steady modeling and three-dimensional time-dependent RMHD modeling. These structures represent characteristic flux tubes anchored within a polar coronal hole, on the edge of an equatorial streamer, and neighboring a strong closed-field active region. We show that the time-averaged properties of the higher-dimensional BRAID models agree well with that of the less-computationally-expensive ZEPHYR models. 

We looked in detail at the energy partitioning, as BRAID imposes no assumptions or restrictions. At heights above the transition region, equipartition is shown to work well to describe the results from BRAID, and is assumed in the ZEPHYR algorithm. We also compared the energy partitioning with predictions from non-WKB reflection for a range of frequencies, showing the limitations of the linear method for such predictions. The time-averaged heating rates from BRAID are lower than the phenomenological expression in Equation (5) for the heating rate below the transition region, rising sharply toward the upper boundary.

In BRAID, Alfv\'en waves are generated by random footpoint motions, whose properties and driving modes are described in Section 2.2. As the Alfv\'en waves propagate upward from the photosphere to the open upper boundary at a height of 2 solar radii, they partially reflect and cause turbulence to develop. With the time-dependence included in BRAID, we are able to show the bursty nature of turbulent heating by Alfv\'en waves. We show that this heating brings energy up into the corona and provide a statistical distribution of energy per event. The more energetic events (i.e., boxes with the most energy) fall within expected nanoflare values.

Overall, we show that time-steady modeling does a good job of predicting the time-averaged results from time-dependent modeling. There is, however, a bounty of information that can be found only by looking at changes in the heating rate over time. Moving from one dimension to three allows the model to contain more realistic physics. We have shown that these models of typical magnetic field structures provide additional compelling evidence to support the idea that Alfv\'en-wave-driven turbulence heats the corona and accelerates the solar wind.

\section*{Acknowledgments}
This material is based upon work supported by the National Science Foundation Graduate Research Fellowship under Grant No. DGE-1144152 and by the NSF SHINE program under Grant No. AGS-1259519. L.N.W. also thanks the Harvard Astronomy Department for the student travel grant and Loomis fund. The authors gratefully acknowledge Adriaan van Ballegooijen for many keen insights and for the development of the BRAID model. The authors also thank the anonymous reviewer for their helpful comments and suggestions.
\vspace{4mm}

\bibliography{ApJ99429R2_arxiv_lwoolsey}
\end{document}